\documentclass[12pt,preprint]{aastex}

\newcommand{\tme}{M$_{\oplus}$}
\newcommand{\mmsun}{\hbox{M}_{\odot}}

\def\la{\mathrel{\hbox{\rlap{\hbox{\lower4pt\hbox{$\sim$}}}\hbox{$<$}}}}
\def\ga{\mathrel{\hbox{\rlap{\hbox{\lower4pt\hbox{$\sim$}}}\hbox{$>$}}}}

\slugcomment{Icarus 161, 431-455}
\shorttitle{Oligarchic growth}
\shortauthors{Thommes et al.}
\begin{document}
\title{Oligarchic growth of giant planets}
\author{E. W. Thommes}
\affil{Astronomy Department, 601 Campbell Hall, University of
California, Berkeley, California 94720}
   
\email{thommes@astro.berkeley.edu}

\author{M. J. Duncan}
\affil{Physics Department, Queen's University, Kingston, Ontario K7L 3N6}

\author{H. F. Levison}
\affil{Southwest Research Institute, 1050 Walnut Street, Suite 426, Boulder,
Colorado 80302}

\begin{abstract}
Runaway growth ends when the largest protoplanets dominate the
dynamics of the planetesimal disk; the subsequent self-limiting
accretion mode is referred to as ``oligarchic growth.''  Here, we
begin by expanding on the existing analytic model of the oligarchic
growth regime.  From this, we derive global estimates of the planet
formation rate throughout a protoplanetary disk.  We find that a
relatively high-mass protoplanetary disk ($\sim$ 10$\times$
minimum-mass) is required to produce giant planet core-sized bodies
($\sim$ 10 M$_{\oplus}$) within the lifetime of the nebular gas
($\la$ 10 million years).  However, an implausibly massive disk is
needed to produce even an Earth mass at the orbit of Uranus by 10
Myrs.  Subsequent accretion without the dissipational effect
of gas is even slower and less efficient.  In the limit of
non-interacting planetesimals, a reasonable-mass disk is unable to
produce bodies the size of the Solar System's two outer giant planets
at their current locations on {\it any} timescale; if collisional
damping of planetesimal random velocities is sufficiently effective,
though, it may be possible for a Uranus/Neptune to form in situ in
less than the age of the Solar System.  We perform numerical
simulations of oligarchic growth with gas, and find that protoplanet
growth rates agree reasonably well with the analytic model as long as
protoplanet masses are well below their estimated final masses.
However, accretion stalls earlier than predicted, so
that the largest final protoplanet masses are smaller than
those given by the model.  Thus the oligarchic growth model, in the
form developed here, appears to provide an upper limit for the
efficiency of giant planet formation.

\end{abstract}


\section{Introduction}
\label{intro}

The initial growth mode in a disk of accreting planetesimals is
runaway growth (eg. Wetherill and Stewart 1989, Kokubo and Ida 1996),
wherein the mass doubling time for the largest bodies is the shortest.
However, when these runaway bodies, or protoplanets, become
sufficiently massive, it is their gravitational scattering (often
called viscous stirring) which dominates the random velocity evolution
of the background planetesimals, rather than the interactions among
the planetesimals.  Since the accretion cross-section of a protoplanet
is smaller among planetesimals with higher random velocities,
protoplanet growth now switches to a slower, self-limiting mode, in
which the mass ratio of any two protoplanets at adjacent locations in
the disk approaches unity over time.  Ida and Makino (1993)
investigated this transition analytically and through N-body
simulations, and Kokubo and Ida (1998, 2000, 2002) studied the
subsequent accretion mode, giving it the name ``oligarchic growth''.
In the terrestrial region, the final accretion phase likely consisted
of the merging of oligarchically-accreted protoplanets;
simulations show that such a process fairly readily produces bodies
with masses comparable to present-day terrestrial planets
(eg. Chambers and Wetherill 1998).  However, an analogous phase in the
trans-Saturnian region would have been highly inefficient; even
sub-Earth mass protoplanets excite each other to high random velocities on
a timescale short compared to their collision timescale, so that only
negligible accretion occurs (Levison and Stewart 2001).  Thus, it
appears that oligarchic growth alone must be called upon to account
for almost all accretional growth in the outer Solar System.

In Section \ref{crossover}, we examine the condition for crossover
from runaway to oligarchic growth, and show that this transition is
expected to set in when the largest bodies are still several orders of
magnitude below an Earth mass.  In Section \ref{growthrate}, we
summarize the previous work on oligarchic growth timescale estimates,
obtain a protoplanet mass function, then extend the model by
considering a system in which the planetesimal surface density varies
in a self-consistent way.  We show that with an approximately tenfold
increase in surface density relative to the minimum-mass model,
protoplanets of mass $\sim$ 10 M$_{\oplus}$ can form.  The standard
nucleated instability model of gas giant formation, wherein a massive
gas envelope accumulates onto a solid core during the nebular gas
lifetime ($\la$ 10 million years, e.g. Strom, Edwards and Skrutskie
1990), is thought to require bodies of this mass (Mizuno et al 1978,
Pollack et al 1996).  Our estimate of the required density enhancement
is somewhat higher than that of Weidenschilling (1998), who finds,
using a multizone statistical simulation, that $4 \times$ the minimum
mass is insufficient to produce giant planet cores, but that an
additional ``modest increase'' is sufficient to make it happen.
 
The formation of ``ice giant'' planets like Uranus and
Neptune at stellocentric distances of $\ga$ 20 AU cannot be similarly
accounted for during this time.  In Section \ref{validity}, we discuss
the validity of the model.  In Section \ref{postgas}, we obtain
oligarchic growth rate estimates in the absence of gas, to ascertain
how much more accretion could have taken place subsequent to the
removal of the nebular gas.
We consider two extremes: that of collisionless planetesimals, and
that of maximally effective collisional damping of random velocities
(though without fragmentation).  In the former limit, Uranus- and
Neptune-mass planets cannot be produced at their current locations on
any timescale unless the initial protoplanetary disk is
implausibly massive; in the latter limit, such planets might be formed
in a reasonable-mass disk and in less than the age of the Solar
System.

In Section \ref{simulations}, we test the semianalytic predictions for
the pre-gas dispersal phase of oligarchic growth against numerical
simulations.  We find good agreement as long as protoplanet masses are
well below their theoretical final masses, however growth in the
simulations stalls early, so that the final masses fall short of those
predicted by the model.  We summarize the results and discuss
implications in Section \ref{discussion}.

\section{Transition to oligarchic growth}
\label{crossover}
Ida and Makino (1993) derive the following condition for the dominance
of protoplanet-planetesimal scattering over planetesimal-planetesimal
scattering in determining the random velocity evolution of the
planetesimal disk:
\begin{equation}
2 \Sigma_M M > \Sigma_m m,
\label{oligarchy_condition}
\end{equation}
where $M$ and $m$ are the protoplanet mass and the effective
planetesimal mass, respectively, $\Sigma_m$ is the surface mass
density of the planetesimal disk, and $\Sigma_M$ is the effective
 surface density of a protoplanet in the disk.  The latter is given by
\begin{equation}
\Sigma_M=\frac{M}{2 \pi a \Delta a_{\rm stir}}
\label{protoplanet surface density}
\end{equation}
where $a$ is the semimajor axis of the protoplanet, and $\Delta
a_{\rm stir}$ is the width of the annulus within which the disk is
gravitationally stirred by the protoplanet.  For a single protoplanet,
this width is about $5.2 a \langle e_m^2 \rangle^{1/2}$, where $\langle
e^2_m \rangle^{1/2}$
is the RMS eccentricity of planetesimals in the disk.  Hereafter, we
shorten the notation to $e_m$, and likewise we write $\langle i_m^2
\rangle^{1/2}$ as $i_m$.  This width is set by the conservation of the
Jacobi energy of planetesimals relative to the protoplanet; details
are given in Ida and Makino (1993).  A simplification inherent in this
approach, and one we shall make use of throughout, is the
representation of the planetesimals as a population of uniform-mass
bodies.  As discussed by Ida and Makino (1993), the effective mass is
the RMS mass of the full distribution.
This requires that $e_m$ and $i_m$ depend only weakly on $m$.  As we
shall see below (Eq. \ref{equilibrium e_m}), this is indeed the case.
At the same time, though, one should bear in mind that the planetesimal mass
spectrum will span many orders of magnitude; as an example, 1 to 100
km radii correspond to a mass range of $10^{-12}$ to $10^{-6}$
M$_{\oplus}$.  The effect of planetesimal sizes on
accretion is considered in Section \ref{effect of size}.

\begin{figure}[p]
\begin{center}
\includegraphics[width=5.0in]{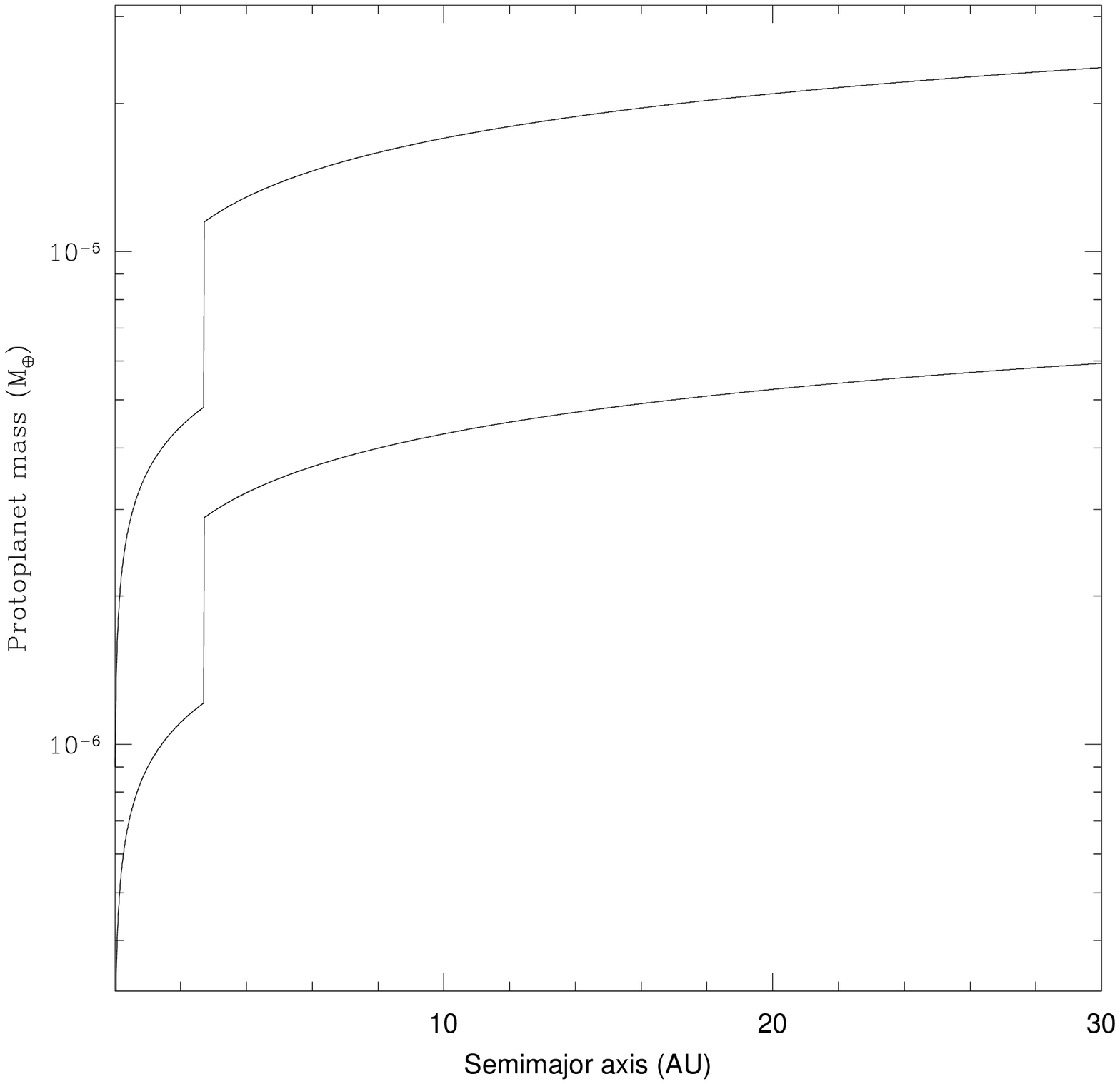}
\caption{Threshold protoplanet mass for crossover from runaway to
oligarchic growth in a minimum-mass (lower curve) and $10\times$
minimum-mass (upper curve) nebula, as obtained from Eq. \ref{crossover_mass}.  A
uniform planetesimal size of 10 km is assumed.}
\label{runaway_stop_minmass}
\end{center}
\end{figure}

When a disk is stirred by multiple protoplanets, one must account for
the overlap of stirred annuli.  One can do this in an average way by
taking
\begin{equation}
\Delta a = \mbox{Min}(5.2 a e, \Delta a_{\rm proto}),
\end{equation} 
where $\Delta a_{\rm proto}$ is the characteristic spacing of protoplanet
orbits.  Kokubo and Ida (1998) find that adjacent protoplanets in a
swarm of planetesimals keep a typical separation of a fixed number $b$
of Hill radii, with $b \sim 10$.  The Hill radius of a body of mass
$M$ and semimajor axis $a$ orbiting a primary of mass $M_*$ is defined
as
\begin{equation}
r_{\rm H} = \left (\frac{M}{3 M_*} \right )^{1/3}a \equiv h_M a,
\label{hillradius}
\end{equation}
where $h_M$ is the reduced Hill radius.  Dispersion-dominated random
velocities means that $e \ga 2 h$, so that $10 r_{\rm H} < 5.2 a e$, and
thus we use $\Delta a = 10 r_{\rm H}$ throughout.  From
Eqs. \ref{oligarchy_condition} and \ref{protoplanet surface density},
the protoplanet mass at which oligarchic growth commences is then
\begin{equation}
\begin{array}{l l l}
M_{\rm oli} & \sim & \displaystyle{\frac{1.6 a^{6/5}b^{3/5}m^{3/5}\Sigma_m^{3/5}}{M_*^{1/5}}}\\
\, & \sim & \displaystyle{ 2.1 \times 10^{-6} \left (\frac{b}{10} \right )^{3/5}
\left ( \frac{M_*}{\rm M_{\odot}}\right )^{-1/5}\left
( \frac{\Sigma_m}{30 {\rm g/cm}^2}\right )^{3/5} \left
( \frac{m}{10^{-9}{\rm M}_{\oplus}} \right )^{3/5} \left
( \frac{a}{{\rm 1 AU}}\right
)^{(6-2 k)/5} {\rm M_{\oplus}} }\\
\end{array}
\label{crossover_mass}
\end{equation}
For the case of an isolated protoplanet ($b \sim 35 - 55$) growing in
a swarm of $m = 10^{23} - 10^{24}$g (200-400 km radius) planetesimals,
Ida and Makino (1993) calculate $M_{\rm oli} \sim 10^{-3} - 10^{-2}$
M$_{\oplus}$.  
Considering a population of planetesimals ($b\sim10$; see above), and
using a perhaps more realistic (eg. Lissauer 1987) planetesimal mass
of $10^{19}$ g ($\sim$10 km) one obtains even lower crossover
masses. Fig. \ref{runaway_stop_minmass} shows $M_{\rm oli}$ throughout the
protoplanetary disk for a one and ten times minimum-mass nebula
(Hayashi 1981; see Eqs. \ref{solids min surface density} and \ref{min
midplane gas volume density} below) with 10 km planetesimals.
$M_{\rm oli}$ is a few $\times 10^{-5}$ M$_{\oplus}$ or less.  Thus,
although runaway growth is much more rapid than oligarchic growth, it
ceases long before protoplanets approaching an Earth mass can form.
For this reason, the contribution to a planet's formation timescale
from oligarchic growth is much more important than that from runaway
growth, and we neglect the latter in our analysis.  
  
\section{Oligarchic growth rate estimates}
\label{growthrate}
When planetesimal random velocities are dispersion-dominated rather than
shear-dominated, the mass accretion rate of an embedded protoplanet is
well described by the particle-in-a-box approximation (Safronov 1969,
Wetherill 1980, Ida and Nakazawa 1989):
\begin{equation}
\frac{d M} {d t} \simeq F \frac{\Sigma_m}{h} \pi R_M^2 \left
(1+\frac{v_{\rm esc}^2}{v_{\rm rel}^2}  \right ) v_{\rm rel},
\label{growthrate1}
\end{equation}
where $h$ the disk scale height, $R_M$ the protoplanet radius, $v_{\rm
esc}$ the escape velocity from the protoplanet's surface, and $v_{\rm
rel}$ the characteristic relative velocity between the protoplanet and
the planetesimals.  $F$ is a factor which compenstates for the
underestimation of the growth rate which results from using the RMS
value of the planetesimal random velocity for $v_{\rm rel}$; its value
is $\sim$ 3 (Greenzweig and Lissauer 1992).  We apply the following
approximations: $h \simeq a i_m$, $i_m \simeq e_m/2$, $v_{\rm
esc}^2/v_{\rm rel}^2 \gg 1$, and $v_{\rm rel} \simeq e_m a \Omega$,
where $\Omega$ is the Keplerian frequency.  These are valid if
gravitational focusing and dynamical friction are both effective; see
for example Kokubo and Ida (1996) for details.  The above equation can
then be rewritten as
\begin{equation}
\frac{d M} {d t} \simeq \frac{C \Sigma_m M^{4/3}}{e_m^2
a^{1/2}} 
\label{growthrate2}
\end{equation}
with $C = 6 \pi^{2/3} [3/(4 \rho_M)]^{1/3} [G/M_*]^{1/2}$, where
$\rho_M$ is the bulk density of a protoplanet.

The planetesimals attain an equilibrium RMS eccentricity, $e_m^{\rm eq}$,
when gravitational perturbations due to the protoplanets are balanced
by dissipation due to gas drag.  Following Ida and Makino (1993), we
obtain $e_m^{\rm eq}$ by equating the viscous stirring timescale due to a
protoplanet of mass $M$, $T_{\rm VS}^{M-m}$, with the eccentricity damping
timescale due to gas drag for planetesimals of characteristic mass
$m$, $T_{\rm gas}^{e_m}$. The former is given by
\begin{equation}
T_{\rm VS}^{M-m} \simeq \frac{1}{40} \left ( \frac{\Omega^2 a^3}{G M} \right
)^2 \frac {e_m^4} {n_{sM} a^2 \Omega}
\label{protoplanet_stirring_timescale}
\end{equation}
(Ida 1990, Ida and Makino 1993), where $n_{sM}$ is the surface number
density of protoplanets, $\Sigma_M/M$.

The gas drag eccentricity damping timescale is given by
\begin{equation}
T_{\rm gas}^{e_m} \simeq \frac{1}{e_m} \frac{m}{(C_{\rm D}/2) \pi r_m^2 \rho_{\rm gas} a \Omega} \equiv \frac{T_{\rm gas}}{e_m}
\label{gas drag timescale}
\end{equation}
(Adachi, Hayashi and Nakazawa 1976).  $C_{\rm D}$ is a dimensionless drag
coefficient, of order 1 for spherical planetesimals of mass $10^{18} -
10^{24}$ g , $r_m$ is the radius of a planetesimal, and $\rho_{\rm gas}$
is the gas volume density.  

Setting $T_{\rm VS}^{M-m} = T_{\rm gas}^{e_m}$ and solving for $e_m$, one obtains 
\begin{equation}
\begin{array}{l l}
e_m^{\rm eq} ( \simeq 2 i_m^{\rm eq} )  & \simeq \displaystyle{ \frac{1.7 m^{1/15} M^{1/3} 
\rho_m^{2/15}}{b^{1/5}C_{\rm D}^{1/5}\rho_{\rm gas}^{1/5} M_*^{1/3}
a^{1/5}} }\\
\,& \displaystyle{ \simeq 0.04 \left ( \frac{b}{10}\right )^{-1/5} \left
( \frac{M_*}{M_{\odot}}\right )^{-1/3} \left ( \frac{\rho_{\rm
gas}({\rm1 AU})}{1.4 \times 10^{-9}\rm{ g/cm}^3}\right )^{-1/5}\left ( \frac{m}{10^{-9}{\rm M}_{\oplus}}\right
)^{1/15} }\\
\,&\displaystyle { \times \left ( \frac{a}{1\rm{ AU}}\right )^{(\alpha-1)/5} \left
( \frac{M}{\rm{1 M}_{\oplus}}\right )^{1/3} }  \\
\end{array}
\label{equilibrium e_m}
\end{equation}
where $\rho_m$ is the planetesimal bulk density and the gas density is
a power law, $\rho_{\rm gas} \propto a^{-\alpha}$.

Substituting Eq. \ref{equilibrium e_m} into Eq. \ref{growthrate2}, one
gets an estimate for the oligarchic-regime growth rate:
\begin{equation}
\frac{dM}{dt} \simeq \frac{3.9 b^{2/5} C_{\rm D}^{2/5} G^{1/2}
M_*^{1/6} \rho_{\rm gas}^{2/5} \Sigma_m }{\rho_m^{4/15}
\rho_M^{1/3} a^{1/10} m^{2/15}} M^{2/3}
\label{growthrate chp 3}
\end{equation}

As a check on the validity of using $e_m = e_m^{\rm eq}$ in calculating
the accretion rate of protoplanets, one can compare $T^{e_m}$, the
timescale to reach $e_m^{\rm eq}$ (obtained from Eq. \ref{equilibrium e_m}
and Eq. \ref{protoplanet_stirring_timescale} or \ref{gas drag
timescale}), to the growth timescale,
\begin{equation}
T_{\rm grow} \equiv \frac{M}{dM/dt}.
\label{growth timescale}
\end{equation}
Using power law gas and solids densities, $\rho_{\rm gas} \propto
a^{-\alpha}$ and $\Sigma_m \propto a^{-k}$, and setting $C_{\rm D}$ = 1,
$\rho_m = \rho_M = 1.5$ g/cm$^3$, one obtains
\begin{equation}
\begin{array}{l l}
\displaystyle{\frac{T^{e_m}}{T_{\rm grow}} \simeq} & \displaystyle{10^{-2} \left ( \frac{m}{10^{-9}
M_{\oplus}}\right )^{2/15} \left ( \frac{\Sigma_m(\mbox{1 AU})}{30 \mbox{g/cm}^2}
\right ) \left ( \frac{\rho_{\rm gas}(\mbox{1 AU})}{1.4 \times
10^{-9}\mbox{g/cm}^3} \right )^{-2/5}} \\
\,& \displaystyle{\times \left ( \frac{M}{\mbox{1
M}_{\oplus}} \right )^{-2/3} \left ( \frac{a}{\mbox {1 AU}}\right
)^{(11-4 \alpha)/15-k}}\\
\end{array}
\label{eqem timescale}
\end{equation}
For a nebula of ten times the minimum mass or less (see below), this
is less than one when the protoplanet mass $M$ is a few $\times
10^{-2}$ M$_{\oplus}$ or larger.

\begin{figure}[p]
\begin{center}
\includegraphics[width=5.0in]{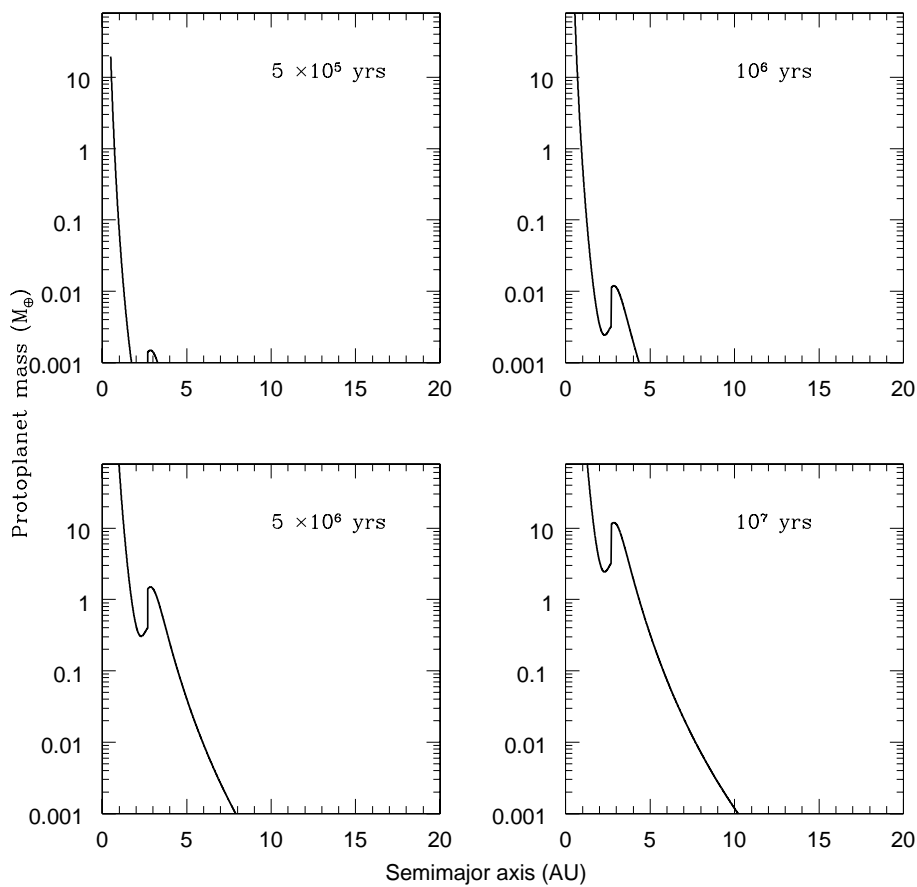}
\caption{The protoplanet mass function for a minimum-mass nebula at
0.5, 1, 5 and 10 Myrs.  The curves are computed using
Eq. \ref{constant surface density mass}, which assumes a planetesimal
surface density that is constant over time, so that the protoplanet mass
(unphysically) increases without bound.  $M_0$ is taken to be zero; as
long as the initial mass is well below the mass range of interest, it
has negligible effect on the solution.}
\label{min_mass_constant_sigma}
\end{center}
\end{figure}

Kokubo and Ida (2000) calculate $T_{\rm grow}$ at the heliocentric
distances of each of the giant planets in the Solar System.  But one
can also directly solve the above differential equation for the
protoplanet mass:
\begin{equation}
\begin{array}{l l l }
M & \simeq& \displaystyle { \left ( \frac{1.3 b^{2/5} C_{\rm D}^{2/5} G^{1/2}
M_*^{1/6} \rho_{\rm gas}^{2/5} \Sigma_m}{\rho_m^{4/15}
\rho_M^{1/3} a^{1/10} m^{2/15}}t + M_0^{1/3}  \right )^3 } \\
\, & \simeq &\displaystyle { \left [ \left (\frac{0.15}{\rm 1 M_{\oplus}} \right )^{1/3} \left (\frac{\rho_{\rm gas}({\rm 1 AU})}{1.4 \times 10^{-9}{\rm g/cm}^2} \right )^{2/5} \left (\frac{m}{\rm 10^{-9} M_{\oplus}} \right )^{-2/15}\left (\frac{a}{\rm 1 AU} \right )^{-1/10-2 \alpha/5-k}   \right . }\\
\,&\,&\displaystyle {\left .\times \left (\frac{\Sigma_m({\rm 1 AU})}{\rm 30 g/cm^2} \right )\left (\frac{t}{\rm 10^5 yrs} \right ) + M_0^{1/3}\right ]^3 },\\
\end{array}
\label{constant surface density mass}
\end{equation}

where we have taken $b$ = 10, $\rho_M$ = $\rho_M$ = 1.5 g/cm$^3$, $C_{\rm D}$=1, and $M_*$ = M$_{\odot}$ in the second line.  
In this way, one obtains an estimate of the mass of protoplanets
throughout the disk at any given time.  Examples are shown
in Fig. \ref{min_mass_constant_sigma}, which plots protoplanet mass
versus semimajor axis at various times.  The calculation is performed
using a surface density of solids based on the minimum-mass model of Hayashi (1981):
\begin{equation}
\begin{array}{c c c c}
\Sigma_{m}^{\rm min} & = & 7.1 (a/\mbox{1 AU})^{-3/2}\mbox{ g/cm}^2, & a < 2.7\mbox{ AU} \\
\,& = & 30 (a/\mbox{1 AU})^{-3/2}\mbox{ g/cm}^2, & a > 2.7\mbox{ AU}\\
\end{array}
\label{solids min surface density}
\end{equation} 
where the discontinuity at 2.7 AU is due to the formation of water ice
at that heliocentric distance (the ``snow line'').  In reality this
boundary may have been significantly less sharply defined.  Likewise,
the snow line's location is quite uncertain; other models place it
around 6 AU (Boss 1995) or 1 AU (Sasselov and Lecar 2000), and it was
likely not stationary as the disk and its temperature profile evolved over time.
In our model disk, we spread the snow line discontinuity over a radial
distance of about 1 AU:
\begin{equation}
\Sigma_m = \left \{ 7.1+(30-7.1)\left [ \frac{1}{2}\tanh \left (\frac{a-2.7
\mbox{AU}}{\rm 0.5 AU} \right )+\frac{1}{2} \right ] \right
\}\left ( \frac{a}{\mbox{1 AU}} \right )^{-3/2}
\label{smoothed minmass disk}
\end{equation}

For the gas component of the disk, we use the midplane value of the
minimum-mass density; this is given by
\begin{equation}
\rho_0^{\rm min}(r)=1.4 \times 10^{-9} (r/\mbox{1 AU})^{-11/4}\mbox{ g/cm}^3,
\label{min midplane gas volume density}
\end{equation}

The full three-dimensional gas density is of the form
\begin{equation}
\rho(r,z)=\rho_0(r) \exp\{-z^2/z_0(r)^2\}\mbox{ g/cm}^3,
\label{min gas volume density}
\end{equation}
where the disk half-thickness, determined by the balance between the
central star's gravity and the gas pressure gradient in the vertical
direction, is
\begin{equation}
z_0(r)=0.0472(r/\mbox{1 AU})^{5/4}\mbox{ AU}
\label{gas half-thickness}
\end{equation}

Fig. \ref{min_mass_constant_sigma} shows a ``front'' of protoplanet
growth sweeping outward through the disk over time.  The front is
quite steep, due principally to the strong dependence of the
protoplanet mass on the plantetesimal surface density (M$\propto
\Sigma_m^3$).  For the same reason, the discontinuity in the surface
density produces a large jump in the protoplanet masses , so that
there is in effect a second, superimposed growth front which is
launched at the snow line.

Since the surface density is taken as time-invariant in this estimate,
the protoplanet mass everywhere in the disk increases without bound
for increasing time.  This is of course unphysical; in reality the
growth of the protoplanets will be constrained by the total amount of
solids in the disk.  At a given location in the disk, the planetesimal
surface density will change over time due to depletion of
planetesimals by accretion, and due to the systematic radial motion of
planetesimals caused by nebular gas drag.  To obtain an estimate of
the former effect, we begin by assuming a series of protoplanets
spaced by $\Delta a =b r_{\rm H}$ (see Section \ref{crossover}).  One can
estimate the planetesimal surface density in an annulus centered on
each one:
\begin{equation}
\begin{array}{l l}
\Sigma_m &\displaystyle{=\Sigma_m^0 - \frac{M}{2 \pi a b r_{\rm H}}}\\
\,&\displaystyle{=\Sigma_m^0 - \frac{(3 M_*)^{1/3}M^{2/3}}{2 b \pi
a^2}},\\
\end{array}
\label{varying surface density}
\end{equation}
where $\Sigma_m^0$ is the original planetesimal surface density at
that location.  Setting $\Sigma_m=0$ also gives us a
limiting protoplanet mass, at which growth stops because the surface
density within $\Delta a$ drops to zero:
\begin{equation}
M_{\rm lim}= \frac{2 \sqrt{2} [b \pi \Sigma_m^0]^{3/2} a^3}{\sqrt{3 M_*}}
\label{mlim}
\end{equation}
This is simply the isolation mass (eg. Lissauer 1987; note that our
$b$ is a factor of two larger than the $B$ in his Eq. 11), modified as
in Kokubo and Ida (1998, 2000) for the case of a population of
planetesimals, so that $b$ is determined by the orbital spacing of
protoplanets rather than by the gravitational reach of a single
protoplanet.

Differentiating Eq. \ref{varying surface density}, one gets the
relationship between the surface density and protoplanet mass rates of change:
\begin{equation}
\left .\frac{\partial \Sigma_m}{\partial t} \right |_{\rm accr}= \frac{-M_*^{1/3}}{3^{2/3} b \pi a^2
M^{1/3}} \frac{dM}{dt}.
\label{sigmarate_accr1}
\end{equation}

In the limit of no planetesimal migration, the behaviour of the system
is obtained by substituting Eq. \ref{varying surface density} into Eq. \ref{growthrate chp 3}:
\begin{equation}
\frac{dM}{dt} \simeq A M^{2/3}(\Sigma_m^0-B M^{2/3})
\label{varying surface density growthrate}
\end{equation}
with
$$
A=\frac{3.9 b^{2/5} C_{\rm D}^{2/5} G^{1/2}
M_*^{1/6} \rho_{\rm gas}^{2/5}}{\rho_m^{4/15}   
\rho_M^{1/3} a^{1/10} m^{2/15}},\,B=\frac{.23
M_*^{1/3}}{b a^2}
$$
Solving this differential equation for the protoplanet mass yields
\begin{equation}
M \simeq \left (\frac{\Sigma_m^0}{B}\right )^{3/2} \tanh \left
[\left ( \frac{1}{3} A {\Sigma_m^0}^{1/2} B^{1/2}\right ) t + \tanh^{-1}(M_0^{1/3} B^{1/2}
{\Sigma_m^0}^{-1/2} ) \right ]^3
\label{varying surface density mass no migration}
\end{equation}

However, to obtain a self-consistent description of the coupled evolution of the
protoplanet masses and the planetesimal disk in the presence of nebular
gas, we must also take into account the planetesimals' orbital
decay.  The rate of change of a planetesimal's semimajor axis under
the action of gas drag is given by
\begin{equation}
\left .\frac{da}{dt} \right |_m \simeq -2 \frac{a}{T_{\rm gas}} \left
(\frac{5}{8}e_m^2+\frac{1}{2}i_m^2+\eta^2 \right )^{1/2} \left
\{ \eta + \left ( \frac{\alpha}{4}+\frac{5}{16}\right )e_m^2 +
\frac{1}{8}i_m^2\right \} \equiv v_m
\label{drag a decay 2}
\end{equation}
(Adachi, Hayashi and Nakazawa 1976).  ${T_{\rm gas}}$ is defined in
Eq. \ref{gas drag timescale} (note that $T_{\rm gas}^{e_m}$ differs from
${T_{\rm gas}}$ by a factor of $1/e_m$), $\alpha$ as before gives the
exponential $a$-dependence of the gas density,
and
\begin{equation}
\eta \equiv \frac{v_{\rm K} -v_{\rm gas}}{v_{\rm K}} =\frac{\pi}{16}(\alpha+\beta)\left
( \frac{c_s}{v_{\rm K}} \right )^2
\label{eta}
\end{equation}
is the fractional difference between the gas velocity $v_{\rm gas}$
and the local Keplerian velocity $v_{\rm K}$ arising from the partial
pressure support of the gas disk; $c_s$ is the sound speed, $\beta$
gives the exponential $a$-dependence of the temperature profile ($T
\propto a^{-\beta}$), and $c_s/v_{\rm K} \simeq z_0(r)/r$.  Since the
rate of orbital decay grows with planetesimal eccentricity and
inclination, one can expect this effect to play an increasingly
important role as protoplanets grow larger (Eq. \ref{equilibrium
e_m}).

Applying continuity, the rate of change of the surface density at a
given radius $a$ in the planetesimal disk due to planetesimal
migration alone is
\begin{equation}
\left . \frac{\partial \Sigma_m}{\partial t} \right |_{\rm migr} =-\frac{1}{a}\frac{\partial}{\partial a}
 \left (a \Sigma_m v_m  \right ).
\label{continuity}
\end{equation}
Using $i_m = e_m/2, e_m = e_m^{\rm eq}$ in Eq. \ref{drag a decay 2}, we
get $v_m$ as a function of $a$ and $M(a)$.  We can then write the
equation of continuity as
\begin{equation}
\left .\frac{\partial \Sigma_m}{\partial t} \right |_{\rm migr} =
-\frac{1}{a} \left [ \Sigma_m v_m + a v_m \frac{\partial
\Sigma_m}{\partial a} + a \Sigma_m \left ( \frac{\partial
v_m}{\partial a}+ \frac{\partial v_m}{\partial M} \frac{\partial
M}{\partial a} \right ) \right ].
\label{continuity 2}
\end{equation}
Implicit in this treatment is the assumption that the net orbital
decay rate of the planetesimals is not significantly affected by the
presence of the protoplanets.  Tanaka and Ida (1997) performed
numerical simulations of protoplanets embedded in a swarm of
planetesimals, and found that a protoplanet shepherds planetesimals
outside its orbit, even when mutual perturbations among the
planetesimals are strong enough to prevent resonant trapping.  This
shepherding effect arises due to the (approximate) conservation of the
planetesimals' Jacobi energy relative to the protoplanet.  However,
they also found that once protoplanets are spaced by 15 Hill radii or
less, adjacent protoplanets' contours of constant Jacobi energy
overlap too much and shepherding no longer occurs.  Since $\Delta
a_{\rm proto}$ is closer to 10 $r_{\rm H}$ for a population of
protoplanets embeddeded in a planetesimal disk (see Section
\ref{crossover}), shepherding is unlikely to play much of a role.

The full oligarchic growth model is thus described by a coupled pair
of partial differential equations.  The first is just
Eq. \ref{growthrate chp 3} with $M=M(a,t)$, $\Sigma_m=\Sigma_m(a,t)$,
while the second is the sum of of the two surface density rates of
change:
\begin{equation}
\frac{\partial \Sigma}{\partial t} = \left .\frac{d \Sigma_m}{dt} \right
|_{\rm accr} + \left . \frac{\partial \Sigma_m}{\partial t} \right |_{\rm migr}
\label{full sigma rate PDE}
\end{equation}

\begin{figure}[p]
\begin{center}
\includegraphics[width=5.0in]{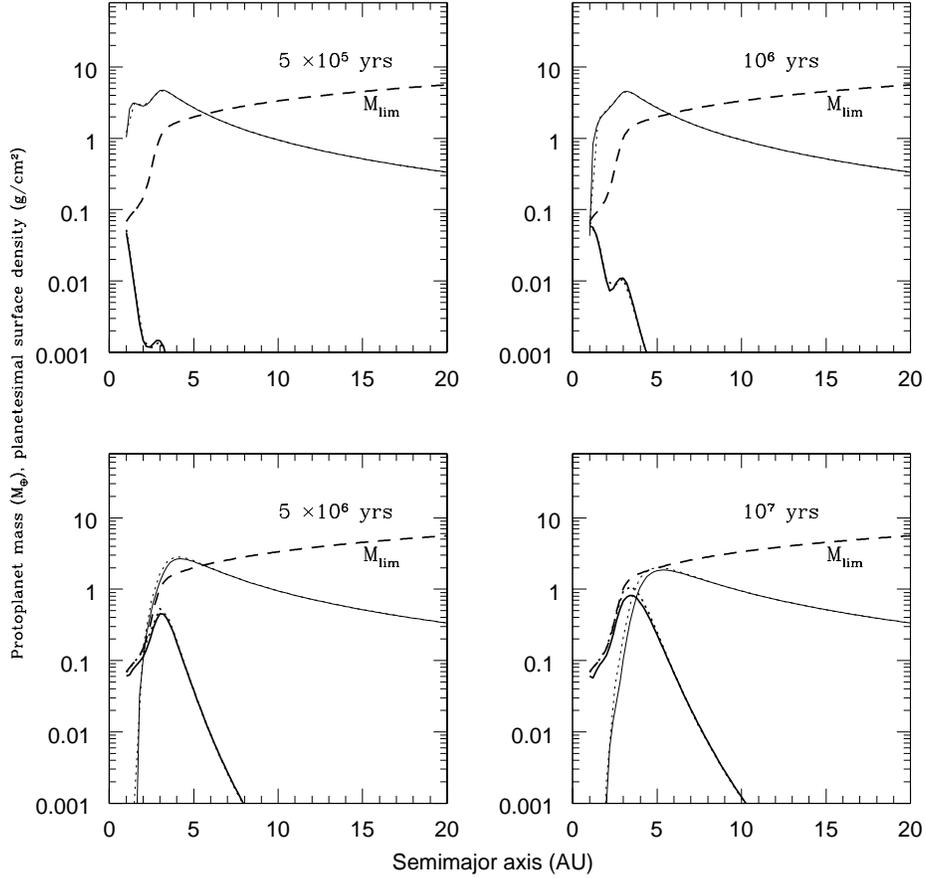}
\caption{Protoplanet mass (thick curves) and planetesimal surface
density (thin curves) versus heliocentric distance for a minimum-mass
gas nebula (Eq. \ref{min midplane gas volume density}) and an
initially minimum-mass planetesimal disk with a smoothed-out snow line
discontinuity (Eq. \ref{smoothed minmass disk}) after 0.5 to 10 Myrs
of evolution.  The solution obtained from numerically integrating the
coupled system of Eqs. \ref{growthrate chp 3} and \ref{full sigma rate
PDE} is shown as solid curves, while the analytic result when migration
of planetesimals is neglected, given by Eqs. \ref{varying surface
density} and \ref{varying surface density mass no migration}, is shown
as dotted curves. The migrationless solution approaches the isolation
mass $M_{\rm lim}$, shown as a thick dashed line, over time.  A
uniform planetesimal radius of 10 km is assumed.
}
\label{min_mass_full_mathematica_solution}
\end{center}
\end{figure}

We solve this system numerically, for an initially minimum-mass disk
identical to the one used to produce
Fig. \ref{min_mass_constant_sigma}.  The evolution of $M$ and
$\Sigma_m$ is plotted in
Fig. \ref{min_mass_full_mathematica_solution}.  Protoplanet growth now
stalls when the planetesimal surface density approaches zero.  The results for
$\Sigma_m$ and $M$ in the limit of no planetesimal migration, given by
Eqs. \ref{varying surface density} and \ref{varying surface density
mass no migration}, are also shown; the latter approaches a limiting
value of $M_{\rm iso}$.  The full solution, however, does not quite
reach $M_{\rm iso}$ because of the depletion of planetesimals by gas
drag orbital decay.
With or without planetesimal migration, by 10 Myrs the largest
protoplanet mass is still less than 1 M$_{\oplus}$.  However, as
mentioned in Section \ref{intro}, the core mass required to form a gas
giant planet by nucleated instability is likely $\sim$ 10
M$_{\oplus}$.  In our own Solar System, measurements of the mass,
radius and gravitational moments of Jupiter and Saturn constrain their
present-day solid core masses to be 0 - 10 M$_{\oplus}$ and 6 - 17
M$_{\oplus}$ respectively (Guillot 1999); it is possible the cores
were originally larger, but have since become partially mixed with
their envelopes (Guillot 2001).  Assuming a protosolar ice/rock ratio,
the core masses of Uranus and Neptune are 85-95\% of their total
masses, that is, 12 - 14 M$_{\oplus}$ and 14 - 16 M$_{\oplus}$
respectively.  This implies that the protoplanetary disk out of which
our Solar System formed was well above the minimum mass, and that a
minimum-mass disk is in fact unlikely to produce any giant planets at
all.

\begin{figure}[p]
\begin{center}
\includegraphics[width=5.0in]{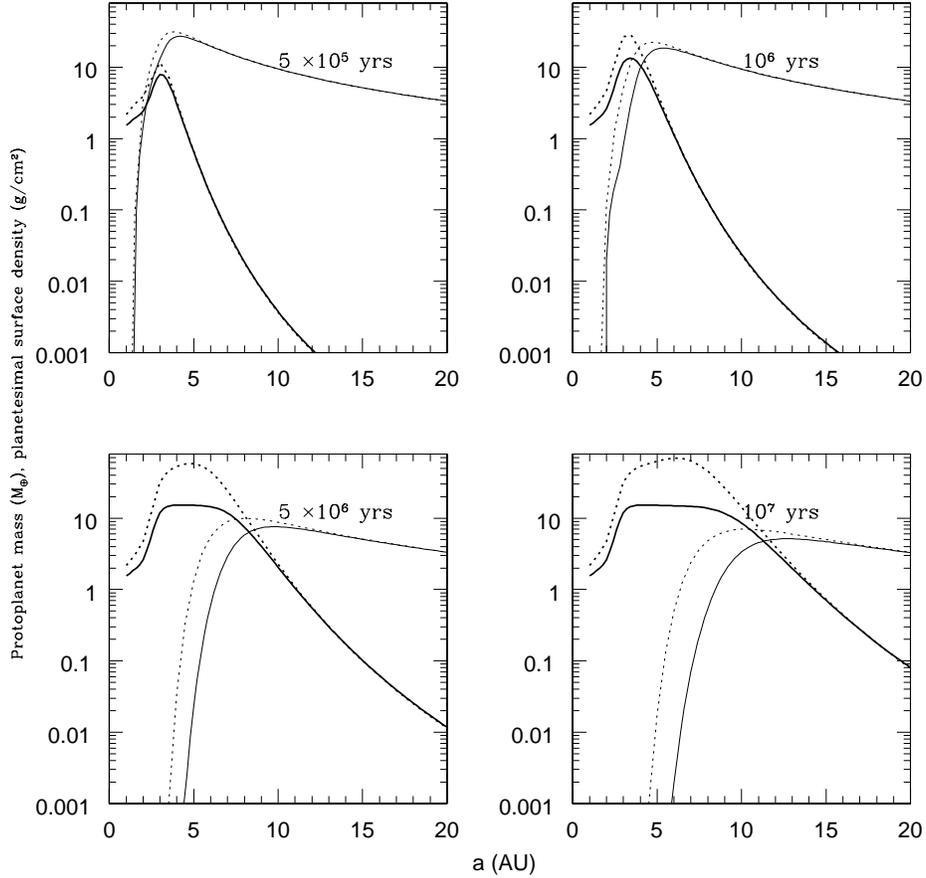}
\caption{Protoplanet mass (thick curves) and planetesimal surface
density (thin curves) versus heliocentric distance for a nebula with
10 times the solids and gas densities of the minimum-mass nebula.
All other parameters are the same as for the minimum-mass case shown
in Fig. \ref{min_mass_full_mathematica_solution}.  The dashed curves
show the migration-less solution computed from Eqs. \ref{varying
surface density} and \ref{varying surface density mass no migration},
while the solid curves show the solution including planetesimal
migration, obtained from numerically integrating Eqs. \ref{growthrate
chp 3} and \ref{full sigma rate PDE}.}
\label{7.5min_mass_nonflat_full_mathematica_solution}
\end{center}
\end{figure}

Fig. \ref{7.5min_mass_nonflat_full_mathematica_solution} shows the
calculation repeated for a disk which
has its solids and gas
densities increased by a factor of ten relative to the minimum-mass
disk.  This yields a solids density of 
about 25 g/cm$^2$ around 5 AU, which is a litte below the
estimate of Lissauer (1987) 
for the density needed to accrete Jupiter's solid core before the
dispersal of the nebular gas.
In the full numerical solution including planetesimal migration, bodies as large as 10 M$_{\oplus}$
form just beyond the snow line in less than 1 Myr.  
Interior to the snow line, the bodies which have formed by this time
are about an order of magnitude smaller.
These results show that a strong dependence of protoplanet formation
rate on initial solids surface density exists even when the depletion
of the disk is accounted for, thus lending support to the theory
that the solids density enhancement beyond the snow line facilitates
giant planet core formation (Morfill 1985).  

In this case, the gas drag-induced orbital decay of planetesimals 
makes a significant difference in the later stages of accretion.  In
the migration-less solution (dashed curves),
protoplanet
masses get above 60 M$_{\oplus}$!  In the full numerical
solution, on the other hand, growth already stalls at a little
over 10 M$_{\oplus}$ due to the rapid orbital decay of planetesimals
stirred by protoplanets of this size.

Also noteworthy is the fact that very little accretion is predicted to
take place in the trans-Saturnian region in 10 Myrs.  By this time, a
$10 \times$ minimum-mass system does not produce even 1 M$_{\oplus}$
bodies beyond about 15 AU.  The protoplanet mass at the location of
Uranus, $\sim$ 20 AU, is only about a tenth of an Earth masses.  The
simplest description of oligarchic growth, in which the planetesimal
surface density is constant (Eq. \ref{constant surface density mass}),
is a good approximation of the early stages of the process, when
protoplanets are small compared to their isolation mass
(Eq. \ref{mlim}).  From Eq. \ref{constant surface density mass}, the
protoplanet mass at a given location and time scales as $\Sigma_m^3$.
Thus the surface density would have to be higher by at the very least
a factor of $(10/0.02)^{1/3}=8$, i.e. 80$\times$ the minimum mass, in
order to produce a body of order 10 M$_{\oplus}$ at the location of
Uranus in 10 Myrs.  The minimum-mass protoplanetary disk contains a
total mass of about 0.02 M$_{\odot}$ within 100 AU; 80 times that is
1.6 M$_{\odot}$, far above the typical range of observationally
inferred disk masses, $\sim$ 0.01 to 0.2 M$_{\odot}$ (eg. Beckwith and
Sargent 1996, Chiang et al 2001).  Also, such a disk would have a
Toomre Q parameter less than one beyond about 20 AU, thus being
unstable to collapse outside that radius, and at best marginally
stable throughout most of the region inside (eg. Nelson et al 1998).

The apparent inability of much of anything to accrete over 10 Myrs in
the trans-Saturnian region presents a definite obstacle to
understanding the formation of our Solar System.  After all, the 1 - 3
M$_{\oplus}$ gas atmospheres of Uranus and Neptune imply that they are
composed of protoplanets which, {\it during} the lifetime of the
nebular gas, grew large enough to retain significant atmospheres.  In
fact, the problem of Uranus and Neptune's formation timescale is even
more severe, as will be discussed in Section \ref{postgas}.

\section{The effect of planetesimal size}
\label{effect of size}

The efficiency of protoplanet accretion in this model is subject to
two competing effects, both of which, for a given nebular gas density,
depend on the characteristic planetesimal size.  On the one hand,
smaller planetesimals experience stronger damping of random
velocities, forming a thinner disk and thus increasing the accretion
rate.  On the other hand, smaller planetesimals are also subject to
faster orbital decay, which depletes the planetesimal surface density
at a given location in the disk more rapidly.  In other words, a
smaller planetesimal size means faster accretion, but also an earlier
end to accretion.  Thus for a finite available time---the lifetime of
the nebular gas---there will be an optimal planetesimal radius $r_m^{\rm
crit}$, which produces the largest final protoplanet mass.  Using,
again, the $10\times$ minimum-mass disk, we compute the protoplanet
mass after 10 Myrs as a function of planetesimal size at different
stellocentric distances.  The result is shown in
Fig. \ref{optimal_mass}.  The optimal protoplanet size at 5, 10 and 30
AU is just under 100 km, a few kilometers, and a few tens of meters,
respectively (one should be a bit cautious about the last value, which
is comparable to the mean free path of gas molecules at 30 AU for our
assumed nebula; this puts us at the limit of the validity of the
Stokes drag law).  Our choice of 10 km planetesimals places us
logarithmically near the middle of the optimal range for the
Jupiter-Saturn region; at the same time, even if the planetesimal size
at 30 AU had happened to be optimal, the largest protoplanet produced
there in 10 Myrs would still have been only a few tenths of an Earth
mass.  Another way to look at the above results is in terms of the
planetesimal size spectrum which will exist in real life (recall that
we are approximating the planetesimals as a uniform-sized population):
Accretion is fastest but least efficient from the small end of the
spectrum, and slowest but most efficient from the large end.

\section{Validity of the estimate}
\label{validity}

A number of simplifications underlie this estimate of protoplanet
growth rates.  To begin with, interactions among planetesimals are neglected
altogether in our analysis.  This seems reasonable since, by
definition, scattering by protoplanets dominates the planetesimal
velocity distribution in the oligarchic regime.  Also, Kokubo and Ida
(2002) showed that the timescale for spreading of the planetesimal
disk due to mutual interactions is large compared to the accretion
timescale (though the disk {\it does} spread significantly due to
stirring by protoplanets, as will be demonstrated by the numerical
simulations in Section \ref{simulations} below).  However, the
protoplanets' dominance of the planetesimal dynamics also means that
the relative velocities among planetesimals are large compared to
their surface escape velocities.  Thus, reprocessing of the
planetesimal population through physical collisions can potentially
bring about a significant decrease in the characteristic planetesimal
size.  Inaba and Wetherill (2001) performed statistical simulations in
the Jupiter-Saturn region; for their adopted fragmentation model, they
found that a substantial fragmentation tail of small bodies formed,
resulting in a high loss rate of solids due to gas drag and
significantly reduced accretion efficiency relative to the case
without fragmentation.  If fragmentation was indeed very effective, it
would constitute a significant obstacle to the formation of giant
protoplanets; it can easily be seen from Fig. \ref{optimal_mass} that
if the majority of the planetesimal mass ends up in, say, bodies with
radii of order 100 m, then the largest protoplanets to form will only
be a bit over an Earth mass.

Another effect of inter-planetesimal collisions will be to dissipate
some of the energy in planetesimal random motions .  We can easily
estimate the importance of this effect relative to the damping by gas
drag.  The timescale for the latter is given by Eq. \ref{gas drag
timescale}.  For the former effect, the lower limit on the timescale
is just the time between inter-planetesimal collisions.  The collision
rate can be estimated as $n (\pi r_m^2) v_{\rm rel}$, where $n$ is the
volume number density of planetesimals, $n \sim \Sigma_m/(m h) \sim
\Sigma_m/(m a e_m)$, and $v_{\rm rel} \sim e_m a \Omega$.  The
collision timescale can thus be written as
\begin{equation}
T_{\rm coll} \sim \frac{4 r_m \rho_m}{3 \Sigma_m \Omega};
\label{tcoll}
\end{equation}
note that it is independent of random velocity.  Setting $T_{\rm coll} =
T_{\rm gas}^{e_m}$ and solving for $e_m$, one obtains the very simple
expression for the eccentricity at which the two
timescales are equal:
\begin{equation}
e_{\rm coll=gas} \sim \frac{2 \Sigma_m}{C_{\rm D} a \rho_{\rm gas}},
\label{coll=gas}
\end{equation}
which is fully determined (within $C_{\rm D}$, which we take to be
one) by just the gas-to-solids ratio plus the profile of the gas scale
height; keeping these fixed, $e_{\rm coll=gas}$ is unchanged when one
scales the disk mass up or down.  We plot $e_{\rm coll=gas}$ for a
Hayashi-profile disk in Fig. \ref{coll=gas}.  For $e>e_{\rm
coll=gas}$, the gas drag timescale is shorter.  Since the collision
timescale has to be the lower limit to the timescale for random
velocity damping by collisions, what we have plotted here constitutes
the upper limit to the eccentricity at which damping by gas drag comes
to dominate over damping by collisions.  This limit is of order a few
times $10^{-3}$, thus from Eq. \ref{equilibrium e_m},
inter-planetesimal collisions can only dominate early on, when only
small bodies have formed and the disk is still dynamically quite cold.
It is therefore reasonable to neglect collisional damping while gas is
present, though one must still remember that collisions may reduce the
characteristic planetesimal size over time.

Perhaps most problematic is the issue of migration due to resonant
interaction with the gas disk.  Protoplanets embedded in a gas disk
launch density waves at their inner and outer Lindblad resonances, and
as a result experience a positive and negative torque from the inner
and outer parts of the disk, respectively (Goldreich and Tremaine
1980).  The outer torque dominates, resulting in a decay of the
protoplanet's orbit, which has been termed Type I migration (Ward
1997).  The migration rate increases with protoplanet mass until the
protoplanet is large enough to open a gap in the gas disk, and then it
is locked into the slower viscous evolution of the disk (Type II
migration).  However, since protoplanets likely do not form a gap
until they reach a mass of $\sim$ 10 - 100 M$_{\oplus}$, and since
before that, the migration timescale at 5 AU can be as short as a few
times $10^4$ years, this poses a problem---not just for this approach
to estimating growth times, but for our understanding of the formation
of giant planets in general.  It has been proposed that fast inward
migration could actually speed accretion (Ward 1986); this requires
that the protoplanet plough through a pristine, dynamically cold
planetesimal disk on its way in.  However, given the shorter growth
timescales at smaller stellocentric radii, a migrating protoplanet
ought to encounter, instead, the dynamically hot remnant of the part
of the disk which has already formed protoplanets.  Furthermore,
simulations have shown that even in the idealized case, accretion
efficiency is low and the disk has to be enhanced by at least a factor
of five relative to minimum mass in order to allow a protoplanet
starting at 10 AU to reach gap-opening mass before it falls into the
star (Tanaka and Ida 1999).

On the other hand, it should be pointed out that the gravitational
interaction of a protoplanet with a gas disk is still far from well
understood.  For one thing, self-consistent simulations of multiple
non-gap-opening bodies in a gas disk have, to our knowledge, not yet
been performed; it is not clear what happens when a number of bodies
launch density waves in close proximity to each other.  Conceivably,
this could lead to only the inner- and outermost members of a
population of protoplanets being strongly coupled to the disk.  Also,
Papaloizou and Larwood (2000) find that for eccentricities such that
the radial excursion is equal to the scale height of the gas disk, the
direction of orbital migration actually reverses.  Thus,
eccentricity-raising interactions among the protoplanets could in
principal counteract their orbital decay.  However, it is difficult to
see how such high protoplanet eccentricities ($h/r \sim 0.07$ at r=5
AU) could arise, since Type I-regime torques cause eccentricity to
decay on an even shorter timescale than semimajor axis.  Furthermore,
it has been shown (Tanaka, Takeuchi and Ward 2002) that if the density
waves excited by a planet reflect at the outer disk edge, the torque
asymmetry can be weakened and a nonmigrating steady state may be
attainable.  Finally, it is possible that the gap opening mass is
smaller than has been thought thus far (eg. Rafikov 2002).  In any
case, we defer the issue of Type I migration during giant planet
accretion to future work.  For the present, whenever the protoplanet
mass somewhere in the disk approaches an Earth mass, either in our
analytic estimates or in the subsequent simulations, one should recall
that we are dealing with the limit of no disk torques.

\section{Oligarchic growth in the absence of gas}
\label{postgas}

We have established above that collisional damping of random
velocities is of little importance in the oligarchic growth regime
while the gas is present.  However, once the gas is removed, this may
no longer be true.  The issue of post-gas accretion is of particular
interest in the case of our Solar System, since from Section
\ref{growthrate}, the in situ formation of Uranus and Neptune during
the gas lifetime appears to be ruled out.  Assuming the problem of the
ice giants' gas content can be otherwise solved, how well can they
grow after the gas is gone?

Given the limited knowledge which exists about the behaviour of
planetesimals in high-speed collisions---not a problem amenable to
direct laboratory study---
it is
quite uncertain just how effectively these would dissipate
energy, and what role fragmentation would play.  We therefore look at
the two extreme cases: a collisionless planetesimal disk, and a disk
in which the random velocity damping timescale is equal to the
collision timescale.

\subsection{The collisionless case}
\label{collisionless case}

Without any damping, an equilibrium planetesimal random velocity no
longer exists.  It thus becomes necessary to simultaneously solve
differential equations for the growth rate and the evolution of the
planetesimal random velocity, $v_m$.  Following the approach of
Safronov (1969), we estimate the latter as
\begin{equation}
\frac{d v_m^2}{dt} = \frac{v_m^2}{T_{\rm rel}}
\label{vratechandra}
\end{equation}
with $T_{\rm rel}$ being the gravitational relaxation time (eg. Binney and
Tremaine 1987):
\begin{equation}
T_{\rm rel} \simeq \frac{1}{n_M \pi(2 G M/v_m^2)^2 v_m \ln \Lambda},
\label{tchandra}
\end{equation}
where $n_M$ is the volume number density of perturbers (protoplanets),
and $\Lambda$ is approximately the ratio of the maximum encounter
distance (taken to be the disk scale height) to the minimum
non-collisional encounter distance (taken to be the gravitationally
enhanced capture radius of the protoplanet); details are given in
Stewart and Wetherill (1988).  Similar expressions derived for the
general velocity evolution of a planetesimal swarm are only valid at
low random velocities, otherwise the relative velocities between
planetesimals are not properly accounted for (eg. Stewart and Ida
2000).  However, for our simple case of one population of bodies on
circular orbits stirring another population, taking the relative
velocity to be the velocity dispersion of the latter should 
constitute a reasonable (rough) approximation.

The accretion rate is given by Eq. \ref{growthrate1} with
$v_{\rm rel}=v_m$ and $h=v_m/(\sqrt{3} \Omega)$.  Eqs.
\ref{growthrate1} and \ref{vratechandra} are integrated, using a
surface 10 times that of the minimum-mass model.  Since the isolation
mass in the giant planet region of such a massive nebula is about an
order of magnitude higher than the mass we are trying to attain ($\sim
10$ M$_{\oplus}$), and since we are interested in the upper limit of
accretion efficiency, we take $\Sigma_m$ to be constant.  Protoplanets
are taken to be initially Mars-mass (0.1 M$_{\oplus}$, well above the
predicted mass at 20 AU at the time of gas dispersal), and the ratio
of velocity dispersion to local Keplerian velocity ($=e$) is initially
set at $10^{-2}$.  The accretion rate is set to zero once
$\sqrt{v_{\rm K}^2+v_m^2}$ is equal to the escape velocity from the
primary, $\sqrt{2 G M_*/a}$, i.e. $v_m=v_{\rm K}$.  Once random
velocities have been raised this high, the density of planetesimals
will have been substantially depleted by ejection, and accretion is
deemed to have effectively ceased.  This is a rather generous estimate
of the length of the regime over which accretion operates; it can also
be argued (eg. Vityazev and Perchernikova 1991) that the planetesimal
disk ought to already be largely depleted when $v_m$ is only $\sim 0.3
- 0.4$  $v_{\rm K}$.  

\begin{figure}[p]
\begin{center}
\includegraphics[width=5.0in]{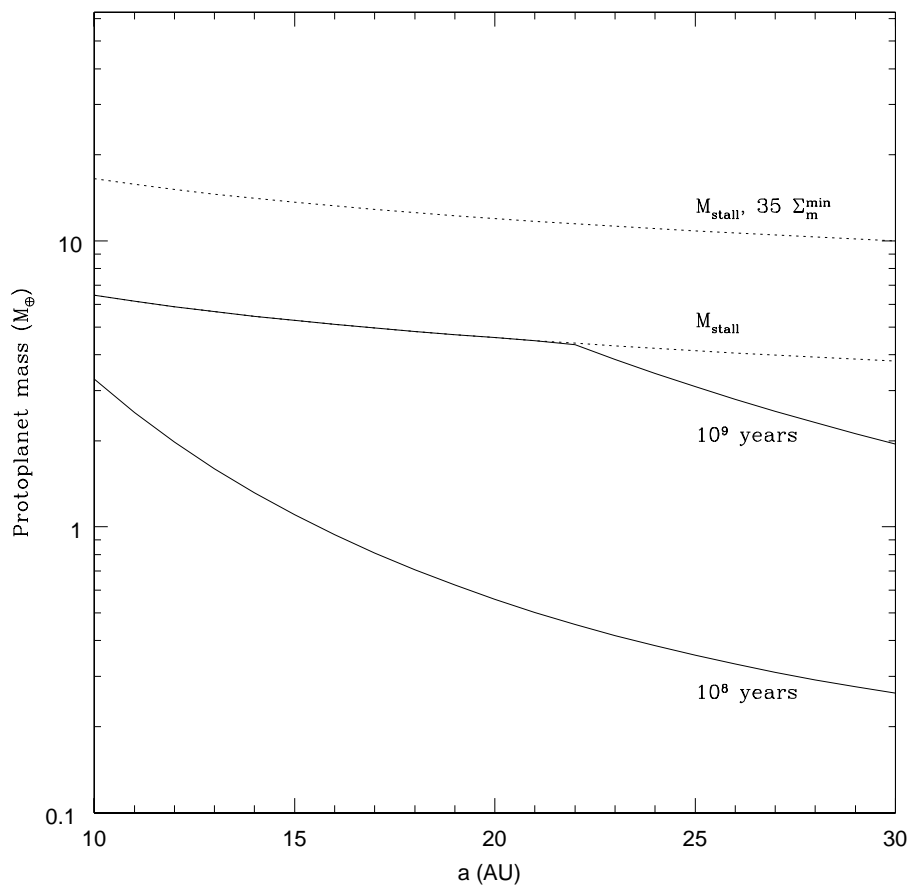}
\caption{Protoplanet mass versus semimajor axis after $10^8$ and
$10^9$ years of post-gas dispersal accretion, in a disk having ten
times the solids surface density of the minimum-mass nebula.  The
lower of the dashed curves shows $M_{\rm stall}$, the protoplanet mass at
which significant escape of planetesimals likely terminates accretion.
The upper dashed curve shows $M_{\rm stall}$ for a 35$\times$
minimum-mass disk.}
\label{postgas_7.5minmass_mathematica_solution}
\end{center}
\end{figure}

Fig. \ref{postgas_7.5minmass_mathematica_solution} shows the results
of the post-gas numerical integration at $10^8$ and $10^9$ years.
Even in $10^9$ years, protoplanets at the locations of Uranus ($\sim$
20 AU) and Neptune ($\sim$ 30 AU) only grow to 4 M$_{\oplus}$ and 2
M$_{\oplus}$, respectively.  Also plotted is $M_{\rm stall}$, the mass at
which accretion is throttled by the escape of planetesimals.  In the
Uranus-Neptune region, it is only 4 to 5 M$_{\oplus}$.  Thus it
appears that in the collisionless limit, a 10$\times$ minimum-mass
disk is unable to oligarchically produce Uranus and Neptune in situ on
{\it any} timescale.
It takes a 35$\times$ minimum-mass disk to make $M_{\rm stall}=10$
M$_{\oplus}$ at 30 AU.  Thirty-five times the minimum mass disk
($\sim$ 0.02 M$_{\odot}$) is about 0.7 M$_{\odot}$.  This is several
times more than the typical range of observed disk masses ($\sim$ 0.01
to 0.2 M$_{\odot}$; see Section \ref{growthrate}).  Also, if one
supposes that the Solar System grew from such a massive protostellar
disk, it becomes difficult to explain why it does not presently
contain more mass---especially in the terrestrial region, where
it is difficult for material to be ejected.

\subsection{The case of perfect collisional damping}
\label{perfect collisions}

Next, we look at what happens when planetesimal random velocities are
damped on the timescale of inter-planetesimal collisions.  In this
case there exists, once again, an equilibrium eccentricity.  It is
obtained by equating the collision timescale, Eq. \ref{tcoll}, to the viscous stirring
timescale, Eq. \ref{protoplanet_stirring_timescale}.  The result is
\begin{equation}
e_m^{\rm eq}=1.7 \frac{m^{1/3} \rho_m^{2/3}}{(b \Sigma_m)^{1/4}} \left
  (\frac{M}{M_{\odot}} \right )^{5/12}
\label{coll equilibrium ecc}
\end{equation}
If we substitute this into the expression for the protoplanet growth
rate, Eq. \ref{growthrate2}, let $\Sigma_m$ be constant in time and
solve, we get
\begin{equation}
M(t) = \frac{D^2}{4} \left (\frac{2 M_0^{1/2}}{D}+t \right )^2
\label{collision protomass}
\end{equation}
where 
$$
D^2 = \frac{18.0 b G M_{\odot}^{2/3} \Sigma_m^3}{(\rho_m \rho_M)^{2/3} m^{1/3} a} 
$$

As mentioned in the previous section, the simplifying assumption of a
constant planetesimal surface density is reasonable as long as
protoplanet masses are far below $M_{\rm lim}$.  In this case, since the
disk is dissipational, the surface density will also decrease due to
planetesimal orbital decay.  However, since we are interested in the
upper limit on protoplanet growth, our approximation will suffice.

\begin{figure}[p]
\begin{center}
\includegraphics[width=5.0in]{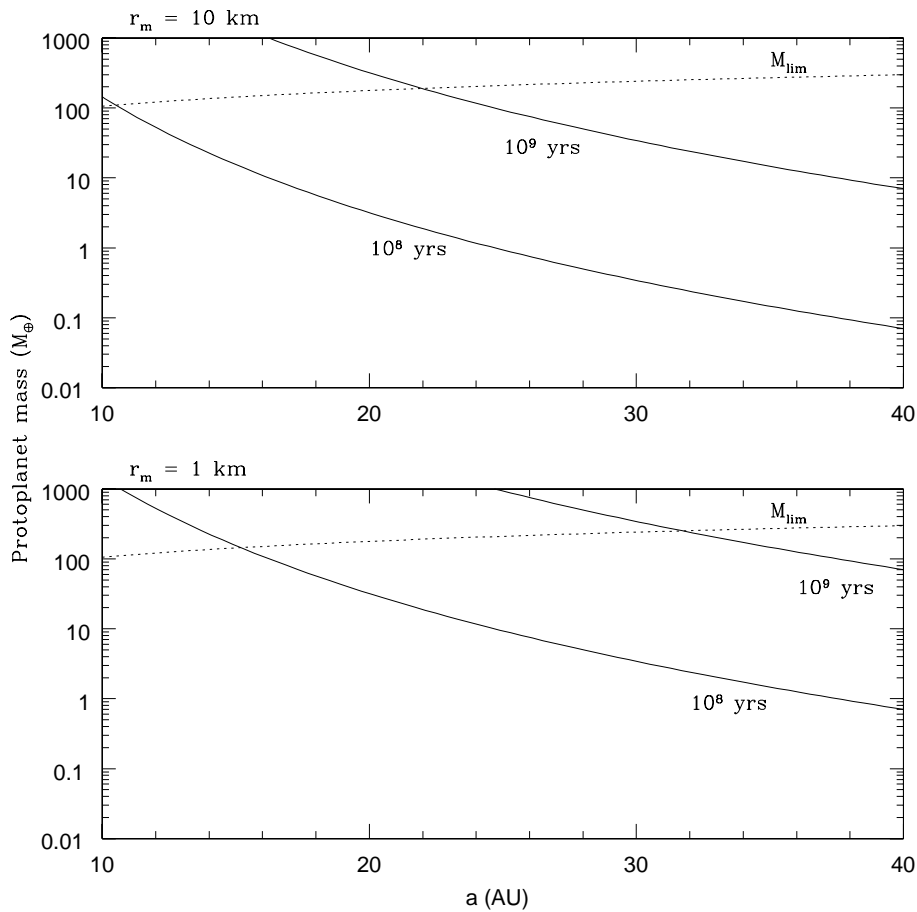}
\caption{The protoplanet mass at $10^8$ and $10^9$ years, in the limit
of maximally effective collisional damping of planetesimal random
velocities.  The dashed curve shows the limiting mass
(Eq. \ref{mlim}), which is simply the mass at which a protoplanet has
accreted all solids in an annulus of width 10 $r_{\rm H}$.  The top panel
shows the results for 10 km planetesimals, while the bottom panel
shows the 1 km case; $M \propto 1/r_m$ so in the latter case the
masses are simply scaled up by a factor of ten.}
\label{protomass_collisions}
\end{center}
\end{figure}

Fig. \ref{protomass_collisions} shows the protoplanet masses computed
from the above expression, for the same 10$\times$ minimum-mass disk
as in the previous section.  The results for both 10 km and 1 km
planetesimals are shown.  In $10^8$ years, the 10 km case produces
neither a Uranus nor a Neptune in situ; the 1 km case does exceed the
mass of Uranus at 20 AU but only reaches a few Earth masses at 30 AU.
By a billion years, both cases have exceeded the masses of Uranus and
Neptune at both 20 and 30 AU.  The protoplanet mass is $\propto
\Sigma_m^3 r_m^{-1} t^2$ as long as $\Sigma_m$ has not changed much
and $M_0$ is small; for example, given that something like a Neptune
can be grown with $\Sigma_m = 10\, \Sigma_m^{\rm min}$ and $r_m=$ 10
km in $t \sim 10^9$ years, one can shorten the timescale to $10^8$ years by
increasing the surface density to a bit under 50 $\Sigma_m^{\rm min}$,
decreasing the planetesimal size to $\sim$ 100 m, or some combination
thereof.  Thus, it would appear that collisional damping does offer a hope for
growing large bodies in the trans-Saturnian region.  However, one must
keep in mind that the above growth rate estimates are very much upper
limits.  For comparison, Davis, Farinella and Weidenschilling (1999)
report on a statistical multizone simulation in the region from 24 to
50 AU, using an approximately 4$\times$ minimum-mass planetesimal
disk.  Though they model collisional damping, growth already stalls at
low masses, and by 4.5 billion years the largest bodies which have
accreted are less than 1 M$_{\oplus}$.

\subsection{The question of Uranus and Neptune}
\label{un}

The actual growth rate after gas dispersal should, at best, lie
somewhere between the limiting cases of Sections \ref{collisionless
case} and \ref{perfect collisions}; even our lower limit may well be
overly optimistic.  Thus, it may not be possible to form Uranus- and
Neptune-like planets at all in the trans-Saturnian region of a
protoplanetary disk, and if it is, the time needed may easily approach
the age of the Solar System.

\begin{figure}[p]
\begin{center}
\includegraphics[width=5.0in]{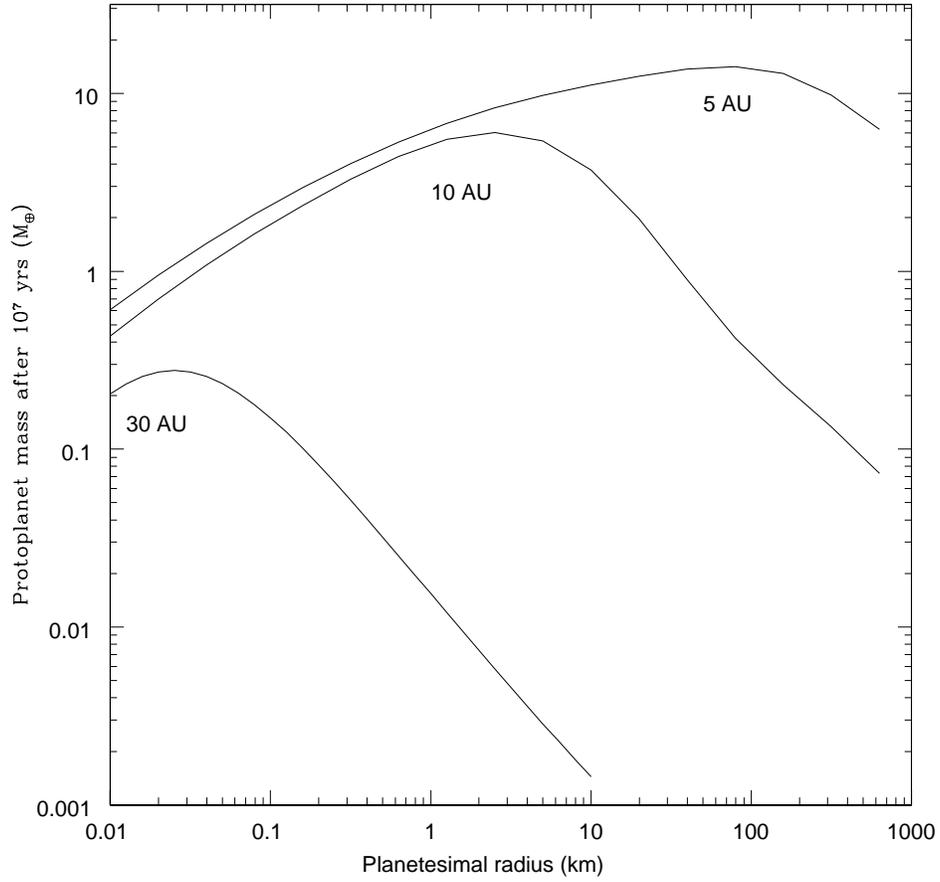}
\caption{The largest protoplanet mass after 10 Myrs of oligarchic
growth as a function of planetesimal radius at 5, 10 and 30 AU, in a 10$\times$ minimum-mass disk.  As can
be seen, the optimal planetesimal radius $r_m^{\rm crit}$ is about 80 km, 2 km and
20 m, respectively.}
\label{optimal_mass}
\end{center}
\end{figure}

One way to shorten the timescale is to presume a larger disk mass, but
as discussed in Section \ref{collisionless case}, one can only go so
far before coming into conflict with observational results.  The other
possibility is to presume that the planetesimal size in the
trans-Saturnian region was very small, either primordially or as a
result of collisional fragmentation, so that collisional damping was
strong.  The primordial size cannot be arbitrarily small, since the
trans-Saturnian region has to survive millions of years of gas drag
without being cleared of planetesimals.  Fig. \ref{optimal_mass} gives
an idea of the lower limit, since if planetesimals are of the optimal
size $r_m^{\rm crit}$ for accretion in the presence of gas (or
smaller), this will leave the region largely cleared of planetesimals
by the time the gas disperses.  Fig. \ref{optimal_mass} shows that
$r_m^{\rm crit}$ is around 20 m at 30 AU, so during the gas-dominated
phase, the characteristic planetesimal size there must have been well
above that in order to allow the possibility of significant post-gas
accretion.  However, planetesimals could have been collisionally
ground down to less than their primordial size subseqent to the
dispersal of the gas.
We defer a more detailed analysis of the role of planetesimal fragmentation the
post-gas oligarchic growth regime---where it may help rather than
hinder accretion---to future work.

An alternative way to account for the existence of Uranus and Neptune
in the outer Solar System is to lift the requirement that they formed
in situ.  Zharkov and Kozenko (1990) propose that during the final
growth phase of a gas giant, it ejects protoplanets outward which can serve
as the starting point for growing the next giant planet.  In this way,
they suggest that Jupiter triggered the formation of Saturn, which in
turn triggered the formation of Uranus and Neptune by launching
outward protoplanets of a few Earth masses.  Simulations by Ipatov
(1991), performed using a Monte Carlo scheme which neglects distant
encounters, suggest that this process works if the protoplanets'
eccentricities remain low throughout.  However, as demonstrated 
above, even after such a head start, the growth of the ice giants
could stall well before they reach their present masses.

The model of Thommes, Duncan and Levison (1999, 2002) assumes that by
the time Jupiter acquired its massive gas envelope, the Jupiter-Saturn
region was able to form, in addition to the solid cores of the gas
giants, two or more extra bodies of comparable mass.  They perform
simulations which show that the mass increase of a giant protoplanet
becoming a gas giant through runaway gas accretion (presumably Jupiter
does this first) causes the remaining protoplanets' orbits to become
unstable.  One or more usually undergo close encounters with
``Jupiter'', and as a result all of them tend to end up on eccentric,
mutually crossing orbits with aphelia in the trans-Saturnian region.
Also included in the simulations is a trans-Saturnian planetesimal
disk, which serves as a source of dynamical friction for the eccentric
protoplanets.  As a result, the protoplanets' eccentricities decrease
over time, decoupling them from Jupiter and from each other on a
timescale of a few million years.  About half of the time, the
simulations produce, after 5 - 10 Myrs, a system which is quite
similar to our own outer Solar System: Two would-be giant planet cores
have ended up on nearly circular, low-inclination orbits with
semimajor axes similar to those of Uranus and Neptune, while the third
is near the present orbit of Saturn.  The timing of the Saturn core's
runaway gas accretion phase is therefore not strongly constrained.
Thus, unlike the picture proposed by Zharkov and Kozenko, Uranus and
Neptune in this model can be transported outward to their present
locations after they have already completed most or all of their
growth.  Also, scattered protoplanets can be recircularized even after
acquiring large eccentricities, by planetesimal disks all the way down
to the minimum mass, in contrast to the findings of Ipatov (1991).

This scenario of outward-scattered ice giants appears to fit well with
the oligarchic growth model, provided the protoplanetary disk is of
sufficient mass to produce the requisite numbers and sizes of
protoplanets.  A disk of 10 times the minimum mass contains 130 Earth
masses of solids between 5 and 10 AU, well above the combined mass of
Uranus and Neptune, plus the upper limits of the heavy element content
of Jupiter and Saturn (Guillot 1999).  As shown in Section
\ref{growthrate}, bodies of mass $\sim$ 10 M$_{\oplus}$ are predicted
to form in less than a million years.  A spacing of $\sim$ 10 $r_{\rm H}$
allows about four such bodies to fit in the Jupiter-Saturn region.
However, the question of how readily this many objects can be produced
in the time available---while both gas and planetesimals are still
present---must ultimately be addressed with numerical simulations.

\section{Numerical simulations}
\label{simulations}

A simple semi-analytic estimate for protoplanet mass as a function of time
throughout a protoplanetary disk is a potentially powerful tool, since
it offers the possibility of characterizing accretional evolution over
time and distance scales which are as yet beyond the reach of
numerical simulation.  Nevertheless, to assess the validity of such an
estimate, comparisons to simulations must be made.  The limits of
computing capacity restrict the domains of full N-body simulations to
relatively narrow annuli within a protoplanetary disk.  In the
simulations presented below, we are able to simulate larger regions of
the disk by making use of simplifications which speed computation but
preserve enough of the relevant physics to make the results
meaningful.

\subsection{Method}

The simulations are performed with a variant of SyMBA, a symplectic
integrator which makes use of an adaptive timestep to resolve close
encounters among bodies (Duncan, Levison and Lee 1998).  This version
also models the aerodynamic drag force on planetesimals due to a gas
disk.  The gas disk is modeled with a three-dimensional density profile of
the form given by Eqs. \ref{min gas volume density} and \ref{gas
half-thickness}, and the parameter $\eta$ is calculated from Eq. \ref{eta}.

In the vertical direction, the
reference Keplerian velocity for the gas disk changes, since it is the
horizontal component of the Solar gravity which provides the central
force: 
\begin{equation}
v_{\rm K}^2(a,z) = \frac{G \mmsun}{a} (a/\sqrt{a^2+z^2})^3
\end{equation}
Stokes drag is applied to planetesimals:
\begin{equation}
\mbox{\boldmath$\dot{v}$} = - K {v_{\rm rel}} \mbox{\boldmath$v$}_{\rm rel},
\label{gasdrag_stokes}
\end{equation}
where $v_{\rm rel}$ is the velocity of a planetesimal relative to the gas disk.
The drag parameter $K$ is
\begin{equation}
K = \frac{3 \rho_{\rm gas} C_{\rm D}}{8 \rho_{m} r_{m}}
\label{drag coeff}
\end{equation}
and we again adopt $C_{\rm D}=1$.

A number of simplifications are made to render feasible the task of
simulating protoplanet growth over a radial range of up to tens of AU
for millions of years.  First, to prevent the computational expense from being prohibitive,
the planetesimal disks are built up of bodies much larger than a
realistic characteristic planetesimal ($\sim$ 1 to 100 km,
corresponding to $10^{-12}$ to $10^{-6}$ \tme).  We adopt planetesimal
masses of 0.01 to 0.05 M$_{\oplus}$.  However, for the purpose of
calculating the gas drag, the actual planetesimal size is not used.
Using the approach of Beauge, Aarseth and Ferraz-Mello (1994) we
instead assume a more physically realistic size of 1 or 10 km ($\sim
10^{-12}$ or $10^{-9}$ M$_{\oplus}$) and apply the drag accordingly.
Thus, each small body in the simulation can be regarded as a
``super-planetesimal'' representing the averaged orbits of a large
number of real planetesimals.

With such large planetesimals, one must ensure that the protoplanets
are sufficiently larger, otherwise they will not experience effective
dynamical friction.  From test runs, we determined that an order of
magnitude difference between the populations is enough to keep the
protoplanets' eccentricities and inclinations reasonably low.  Also,
using super-planetesimals increases the planetesimal accretion
rate of a protoplanet, since the physical interaction radius is the
sum of both bodies' radii, and since in the realistic case the
planetesimal radius is negligibly small compared to that of the
protoplanet.  A factor of ten mass difference gives a factor of about
two difference in radii, and thus a growth rate $\propto
(r_M+r_m)^2$ which is initially too high by a factor of
roughly two.  The fractional error decreases as the protoplanets grow
and widen the protoplanet-planetesimal size gap.

 The large starting mass of the protoplanets constitutes an
unrealistic initial condition; for example, from the semianalytic
estimate, protoplanet masses at 10 AU need several million years to
reach 0.1 M$_{\oplus}$ if they are initially $\ll$ 0.1 M$_{\oplus}$
(Fig. \ref{7.5min_mass_nonflat_full_mathematica_solution}).  Thus the
simulations give the protoplanets a significant head start.  Again,
this leads to a smaller fractional error in protoplanet growth rates
for larger protoplanets, and since the timescale to reach the final
mass is dominated by the time spent in the later phases, the oversized
protoplanets are a reasonable initial condition.

Finally, just as in the semi-analytic estimate,
planetesimal-planetesimal interactions---both gravitational and
collisional---are neglected.  The planetesimals are treated as a
non-self-interacting population by the integrator, though each one
fully interacts with all protoplanets, thus making these
``N+N$^\prime$-body'' simulations in which the computation time scales
quadratically with N (the number of protoplanets) but only linearly
with N$^\prime$ (the number of planetesimals).  Another benefit of
neglecting gravitational interactions among planetesimals is that it
prevents self-stirring of the planetesimal population which, given
their large masses and lack of softening---at present SyMBA does not
support softened potentials due to the way close encounters are
handled---would result in unrealistically high eccentricity and
inclination growth rates.  (Partially) inelastic collisions between
planetesimals, on the other hand, would act to reduce their random
velocities, but as shown in Section \ref{validity}, this effect is
neglible compared to damping by gas drag.
In the simulations, collisions between protoplanets and planetesimals,
or between two protoplanets, are treated as perfectly inelastic; that
is, the two participating bodies are always merged.  Work done on the
role of fragmentation in late-stage planetary formation (Alexander and
Agnor 1998) suggests that this is a reasonable assumption.

\subsection{Simulation results}
\subsubsection{Run A}

The first simulation is performed in the vicinity of the snow line
(2.7 AU) in the Hayashi disk.  The gas and solids surface densities
are increased everywhere by a factor of five relative to the minimum
mass model.  Equal-mass, 0.01 M$_{\oplus}$ planetesimals are initially
distributed between 1.5 and 5 AU.  Planetesimals initially have a
Rayleigh distribution in eccentricities and inclinations, with RMS
values of 0.01 and 0.005 ($=0.29 \degr$) respectively, somewhat lower
than what is given by Eq. \ref{equilibrium e_m} ($e_m \simeq 0.03
\simeq 2 i_m$).  However, since the
timescale to reach $e_m^{\rm eq}$ once the simulation commences is short
compared to that for accretion (see Eq. \ref{eqem timescale}),
unrealistically low initial planetesimal random velocities have
little effect on the outcome of the simulation.  The protoplanets
are given equal masses of 0.1 \tme, and are distributed over the same
range in semimajor axis, with succesive bodies spaced about 10 $r_{\rm H}$
apart.  Planetesimal densities are set at 3 g/cm$^3$.  The density of
ice-enhanced material beyond the snow line would have been
lower---perhaps only half of this---but since growth rates are not a
strong function of the body densities, a single density is used for
the planetesimals.  Protoplanets outside the snow line are given an
initial density of 1.5 g/cm$^2$, but because SyMBA averages densities
when bodies merge, their densities also approach 3 g/cm$^3$ as they
accrete planetesimals.  Gas drag commensurate with a size of 10 km is
applied to the planetesimals.  The fractional difference $\eta$
between Keplerian velocity and the gas orbital velocity
is of order $10^{-3}$.

\begin{figure}[p]
\begin{center}
\includegraphics[width=5.0in]{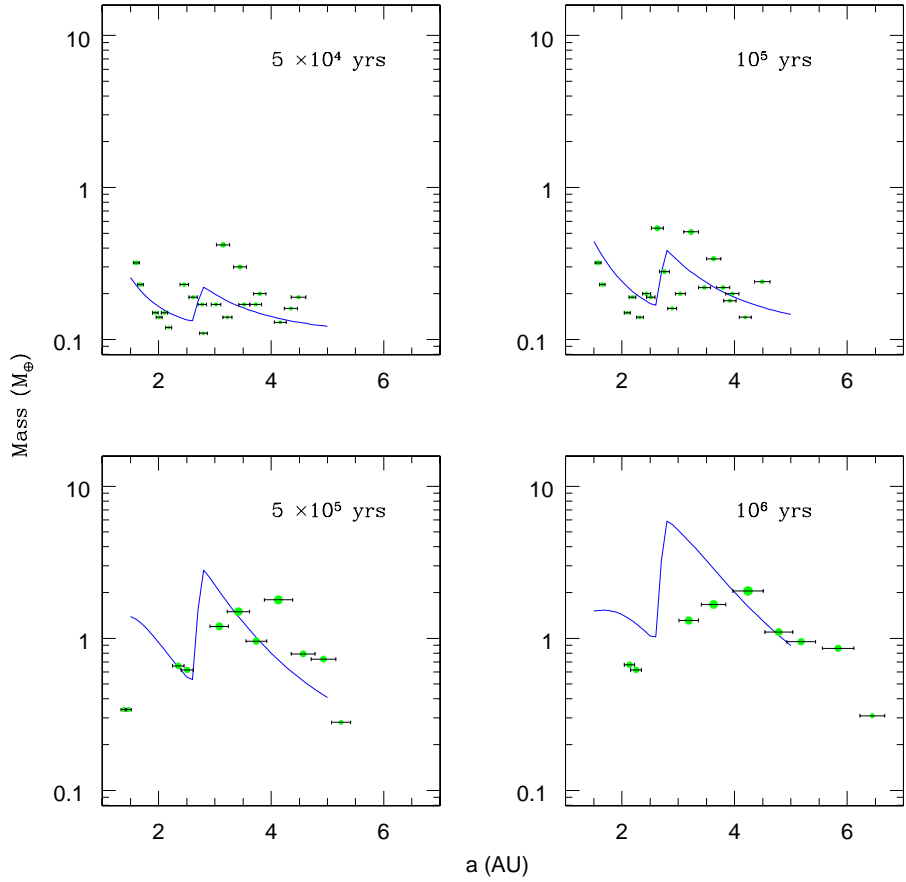}
\caption{Snapshots of the protoplanet masses and semimajor axes in Run
A at different times.  The superimposed curve shows the mass predicted
by the oligarchic growth model at each time, and the horizontal ``error
bars'' on each protoplanet have a length of 10 Hill radii.} 
\label{a_ssdr_massonly}
\end{center}
\end{figure}

The total simulation time is 1 Myr.  Snapshots of the protoplanet
masses and semimajor axes at .05, .1, .5 and 1 Myrs are shown in
Fig. \ref{a_ssdr_massonly}.  Also plotted are the estimates of
protoplanet mass, obtained from solving Eqs. \ref{growthrate chp 3} and
\ref{full sigma rate PDE} with an initial mass of 0.1 M$_{\oplus}$,
and with the physical protoplanet cross-section increased to take into
account the large planetesimal radii ($r_M \rightarrow r_M+r_m$).
Protoplanet growth stalls earlier than predicted, with little change
in protoplanet massess between 0.5 and 1 Myrs; there is primarily just an
overall spreading of protoplanet orbits over that time.  Since the
planetesimals spread with the protoplanets, the expansion of the
protoplanet system beyond the original radial extend of the
planetesimal disk acts to lower the planetesimal surface density, and
thus slow growth, relative to the analytic description.  The largest
protoplanet at 1 Myr has a mass of 2 M$_{\oplus}$; the largest
model-predicted mass at this time is around 6 M$_{\oplus}$.  The assumption of a 10 $r_{\rm H}$
orbital spacing is roughly borne out by the largest protoplanets.

A jump in protoplanet mass does seem to exist at the snow line for a
while, particularly at 0.1 Myrs.  At later times this jump becomes
washed out.  But this is to be expected, since both protoplanets and
planetesimals diffuse in semimajor axis as they gravitationally
interact with each other, and since the simulation does not
``enforce'' the snow line once it starts running.  In other words,
sublimation/freezing of ice/water crossing the boundary is not
modeled.

Fig. \ref{1e6_paper} shows the state of the simulation at 1 Myr in
more detail, including planetesimal eccentricities and inclinations.
The eccentricities and inclinations are comparable to the predicted
values beyond 3 AU.  However, inside 3 AU, only two protoplanets
remain and these have opened a shared gap, thus halting accretion.  As
a result, the planetesimal random velocities here are significantly
lower than predicted.  The effect of migration of planetesimals due to
gas drag can be clearly seen here.  Over time, more and more are
deposited interior to (most of) the protoplanets, forming a
broadening, dense, dynamically cold ring.  This happens because less
eccentric planetesimal orbits decay less rapidly (see Eq. \ref{drag a
decay 2}), so that once planetesimals are out of range of strong
gravitational stirring by the protoplanets, their radial motion slows
and they pile up.  Orbital repulsion between the protoplanets and
these massive rings---at $10^6$ years, there are about 14 M$_{\oplus}$
in planetesimals between 2.5 and 3 AU, and about 16 M$_{\oplus}$
interior to 2 AU--- further promotes the segregation between the
planetesimal and protoplanet populations.

The formation of dynamically cold planetesimal rings like those seen
in this and subsequent runs is a simulation artifact, unlikely to
occur in a real system.  In actuality these planetesimals would have
been further scattered by protoplanets already formed at smaller
stellocentric distances.  Alternatively, in cases where a high density
of planetesimals really did accumulate in a region largely devoid of
gravitational stirring by protoplanets, they would simply revert
temporarily to runaway growth and spawn new protoplanets in their
midst.  Therefore, the end state of Run A is almost certainly a poor
representation of reality interior to 3 AU.  In particular, the dense
planetesimal ring between 2.5 and 3 AU would have produced much more
protoplanet growth, and thus the analytic estimate of protoplanet mass
would have been better reproduced in that region than it is in the
simulation.  At the same time, as long as a region of the disk is, for
whatever reason, predominantly stirred by only one or a few
protoplanets, the opening of a gap like that in Run A (and subsequent
runs) is a likely outcome (Rafikov 2001).

\subsubsection{Run B}
\label{Run B}

The next simulation we present uses a planetesimal disk with an
initial surface density
\begin{equation}
\Sigma_m^{\rm B} = 250 (a/\mbox{1 AU})^{-2}\mbox{g/cm}^2=10(a/\mbox{5 AU})^{-2}\mbox{g/cm}^2, 
\label{solids surface density 1}
\end{equation}
extending from 5 to 15 AU.  This is suggested by Pollack et
al (1996) as the optimum surface density for the accretion of the
giant planets.  It is 3.7 times the minimum-mass surface density at
5 AU.  The corresponding gas nebula is therefore given a midplane
volume density of 3.7 times its value at 5 AU and a density
profile $\propto a^{-2-5/4} = a^{-13/4}$, under the assumption that
the gas scale height also has the profile of Eq. \ref{gas half-thickness}.
The gas volume density is thus
\begin{equation}
\rho_{\rm gas}^{\rm B}=1.15 \times 10^{-8}(a/\mbox{1 AU})^{-13/4}\mbox{g/cm}^3
\label{gas density 1}
\end{equation} 
The planetesimal disk consists of equal mass bodies of 0.02
M$_{\oplus}$.  In this disk, protoplanets of initial mass 0.2
M$_{\oplus}$ are placed at intervals of approximately ten Hill radii.
For the application of gas drag, the planetesimals are, again, assumed
to have a size of 10 km.  

\begin{figure}[p]
\begin{center}
\includegraphics[width=5.0in]{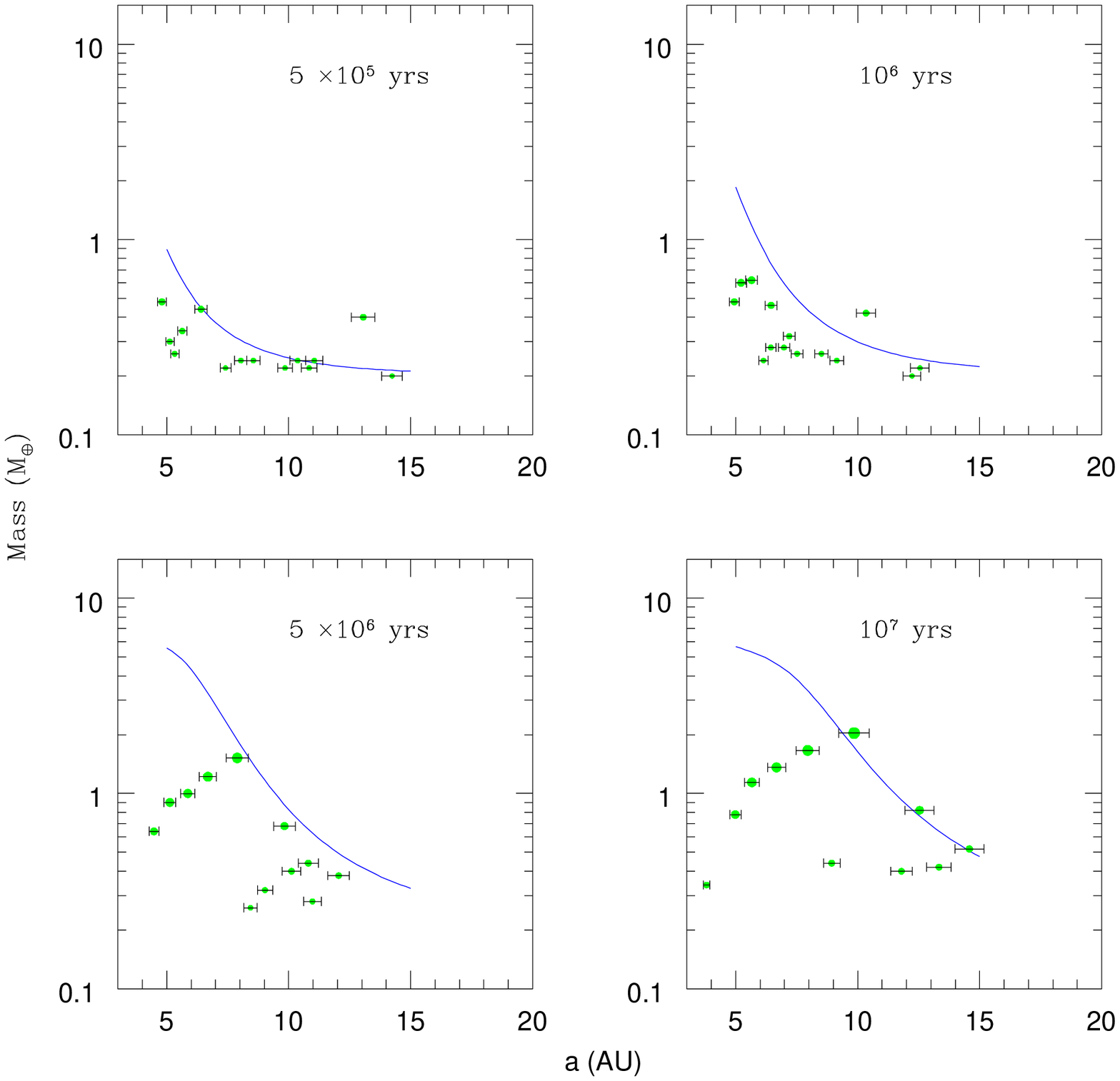}
\caption{Snapshots of the protoplanet masses and semimajor axes in Run
B at different times, together with the model-predicted mass
function.  The horizontal bars on the protoplanets are of length 10 $r_{\rm
H}$.}
\label{a_fixed_drag_massonly}
\end{center}
\end{figure}

\begin{figure}[p]
\begin{center}
\includegraphics[width=5.0in]{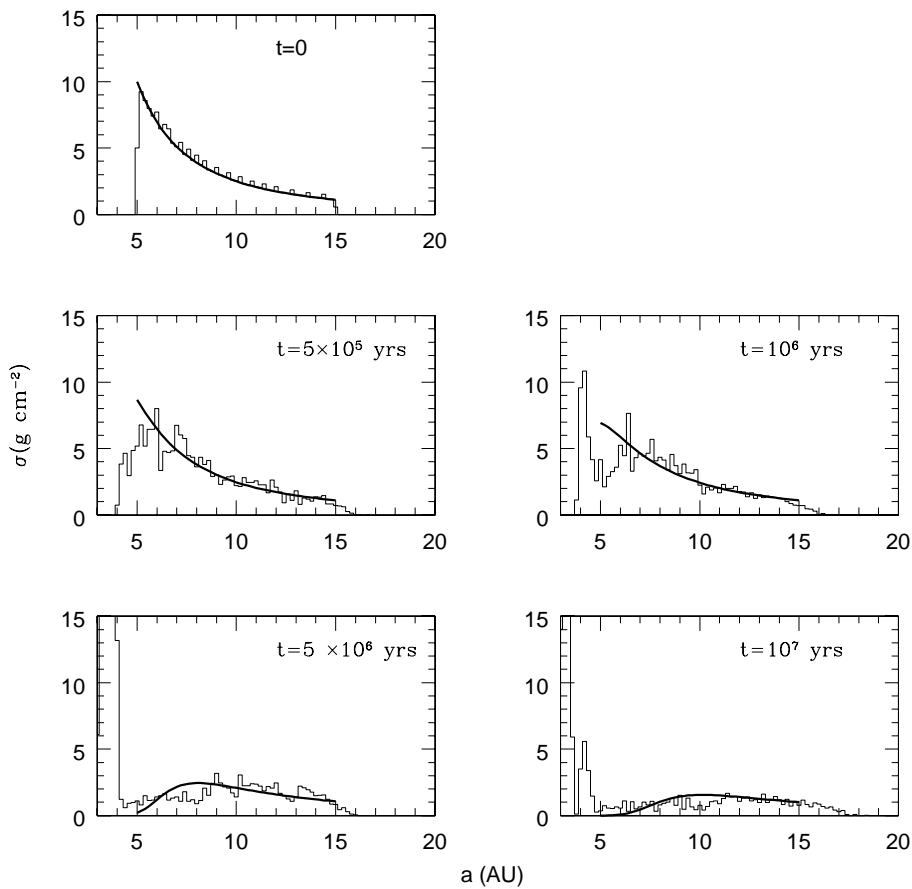}
\caption{Evolution of the planetesimal disk surface density in Run B
(histograms), revealing simulation edge artifacts.  The superimposed
curves show the model-predicted surface densities.  By about a million
years, the planetesimal surface density in the simulation falls well
short of the theoretical value near the inner edge of the protoplanet
population at 5 AU; by 10 Myrs, the underdense region has moved all the way out
to $\sim$ 11 AU.  At the same time, inward-migrating planetesimals
pile up interior to the protoplanets, leading to locally very high
surface densities.}
\label{sigma_snaps}
\end{center}
\end{figure}

The length of the run is 10 Myrs.  Snapshots of the protoplanet masses
and semimajor axes at 0.5, 1, 5 and 10 Myrs are shown in
Fig. \ref{a_fixed_drag_massonly}.  Fig. \ref{1e7_log} gives the state
of the run at 10 Myrs in more detail, showing also the eccentricities
and inclinations of the protoplanets and planetesimals.  The largest
protoplanet which has formed by this time has a mass of 2
M$_{\oplus}$.  The location of the growth front matches that predicted
by the model quite well throughout the run.  Agreement with the
theoretical protoplanet masses beyond 5 AU is likewise quite good
initially, however the masses reached after the growth front has
largely swept past, inside 6 - 10 AU at 1 - 10 Myrs, are smaller than
the predicted final masses by a factor of several.  This appears to be
at least partly an edge effect: Under the action of stirring by
protoplanets, the planetesimal disk spreads---this is superimposed on
the net inward migration of planetesimals---which further lowers the
disk surface density, particularly near the edges.  Since protoplanets
grow fastest and orbital times are shortest at the inner edge, the
surface density there is affected most strongly.
Fig. \ref{sigma_snaps} shows that the surface density near the inner
edge of the protoplanet population, around 5 AU, is indeed lower than
the analytic estimate after about a million years.  As time goes on,
this underdense region moves outward; by 10 Myrs, it reaches all the
way out to about 11 AU.  In addition, Fig. \ref{1e7_log} shows that,
similar to the case of Run A, the innermost protoplanet at 10 Myrs has
cut short its growth by forming a gap in the planetesimal disk.

\begin{figure}[p]
\begin{center}
\includegraphics[width=5.0in]{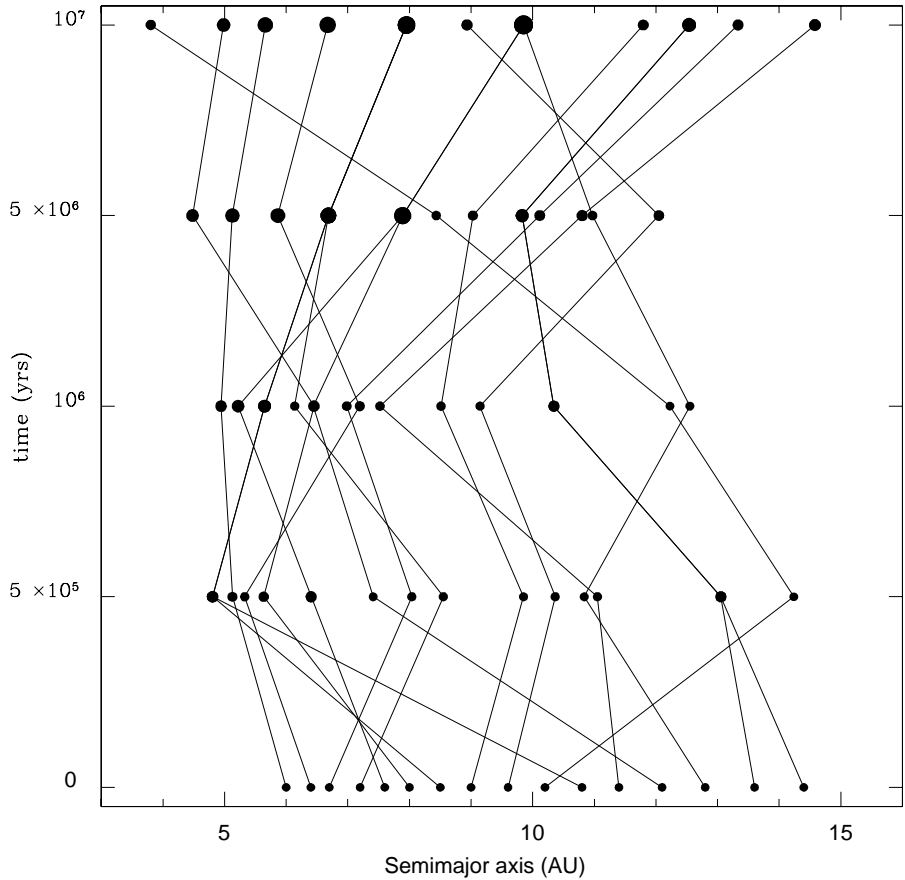}
\caption{Evolution of protoplanet semimajor axes in Run B.  The
plotted circles are proportional to the protoplanets' physical sizes.
The connecting lines trace the history of each protoplanet.  Going
from bottom up, wherever multiple lines go in and one line comes out
of a protoplanet, mergers have occurred in the intervening time; there
are five mergers over the course of this run (note that there are no
mergers at $10^6$ yrs, only two protoplanets with nearly the same
semimajor axis, about 6.5 AU)}.
\label{runb_evol}
\end{center}
\end{figure}

Fig. \ref{runb_evol} shows the evolution of protoplanet semimajor axes
in detail, including their merger history.  A total of five mergers
among protoplanets take place, reducing their number from the original
16 to 11; no protoplanets are lost to ejection, or otherwise leave
simulation domain.  In this run, as in the others, mergers among
protoplanets only play a secondary role in the accretion process;
sweep-up of planetesimals is the primary growth mode, therefore the
oligarchic growth model provides a reasonable fit.  Also apparent in
Fig. \ref{runb_evol} is that significant reordering of protoplanet orbits takes place over the
course of the run.

\subsubsection{Run C}

\begin{figure}[p]
\begin{center}
\includegraphics[width=5.0in]{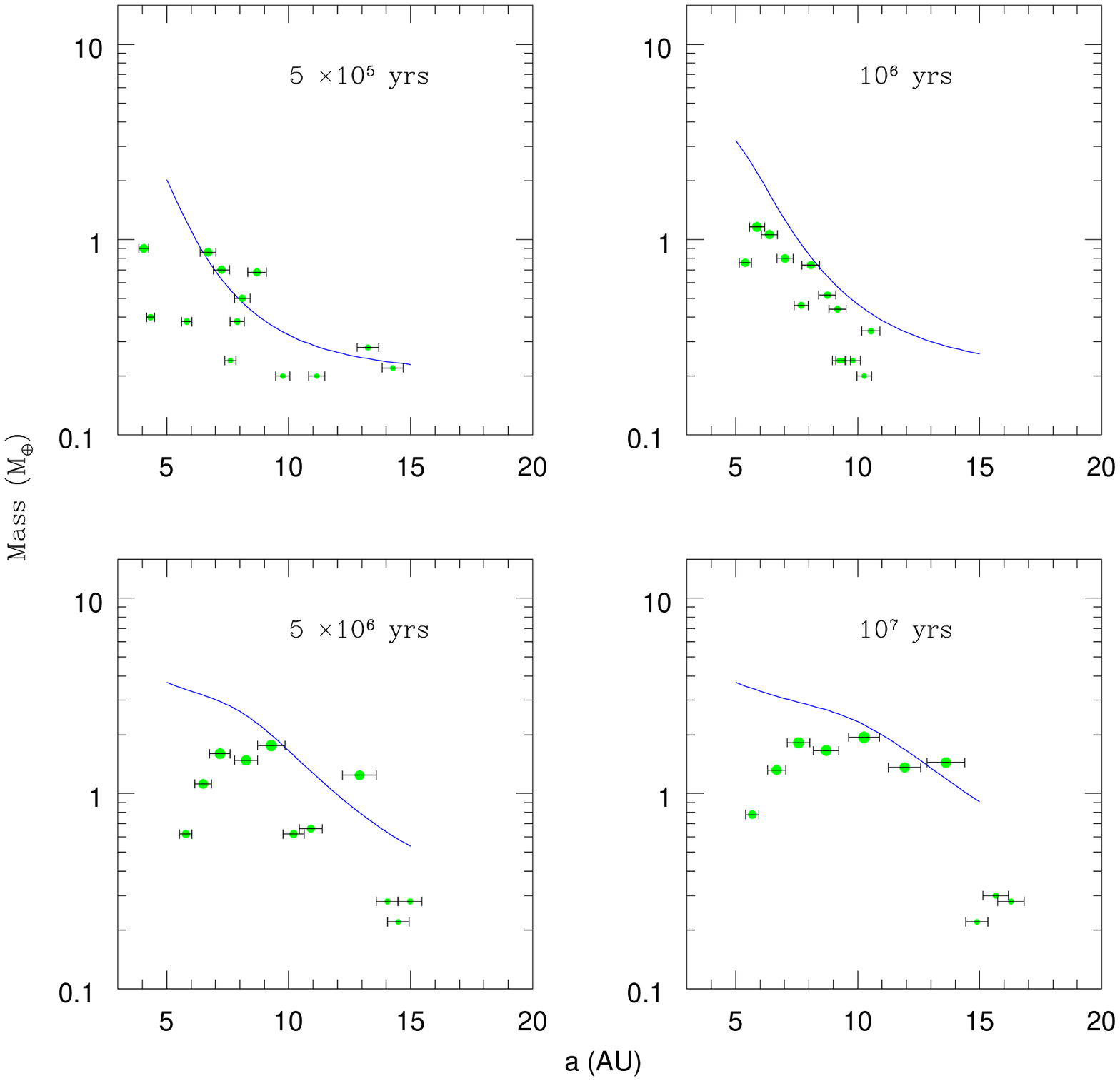}
\caption{Snapshots of the protoplanet masses and semimajor axes in Run
C at different times, together with the model-predicted mass
function.  The horizontal bars on the protoplanets are of length 10 $r_{\rm H}$.}
\label{a_1km_massonly_rev1}
\end{center}
\end{figure}

We repeat the above simulation, but with the drag force on the
planetesimals increased to what 1 km objects would experience.
Snapshots of the run are shown in Fig. \ref{a_1km_massonly_rev1}, and
the endstate, at 10 Myrs, is shown in Fig. \ref{1km_10myrs}.  In
keeping with what the model predicts, the growth front moves outward
more rapidly than in Run B.  The largest protoplanet at 10 Myrs is
just under 2 M$_{\oplus}$, very similar to Run B, although the
predicted largest protoplanet mass is only about half that of Run B.
The theoretically predicted discrepancy comes about because smaller
planetesimals ought to bring about faster growth, at the expense of a
smaller final mass, as discussed in Section \ref{effect of size}.
However, the analytic protoplanet masses in Run B are only larger than
those of Run C inside about 8 AU at 10 Myrs.  In both runs, the growth
of protoplanets in the inner region is stunted by simulation edge effects,
as described in Section \ref{Run B} above.
 
\subsubsection{Run D}

Lastly, we show the result of a simulation performed with a more
massive initial protoplanetary disk.  The solids surface density is
chosen as
\begin{equation}
\Sigma_{m}^{\rm D} = 20 (a/5\,AU)^{-3/2} \rm{ g/cm}^3, 
\label{sigma_d}
\end{equation}
which is about 7.5 times that of the minimum-mass nebula.
The gas surface density is
increased above minimum mass by the same factor, and the
half-thickness is assumed to be, again, given by Eq. \ref{gas
half-thickness}.  The total mass in planetesimals is about 2.7 times
as large as in Run B, so to keep the number of bodies reasonably low,
more massive bodies are used in Run D.  The planetetesimal mass is
changed to 0.05 M$_{\oplus}$, and the protoplanet initial mass to 0.5
M$_{\oplus}$.

\begin{figure}[p]
\begin{center}
\includegraphics[width=5.0in]{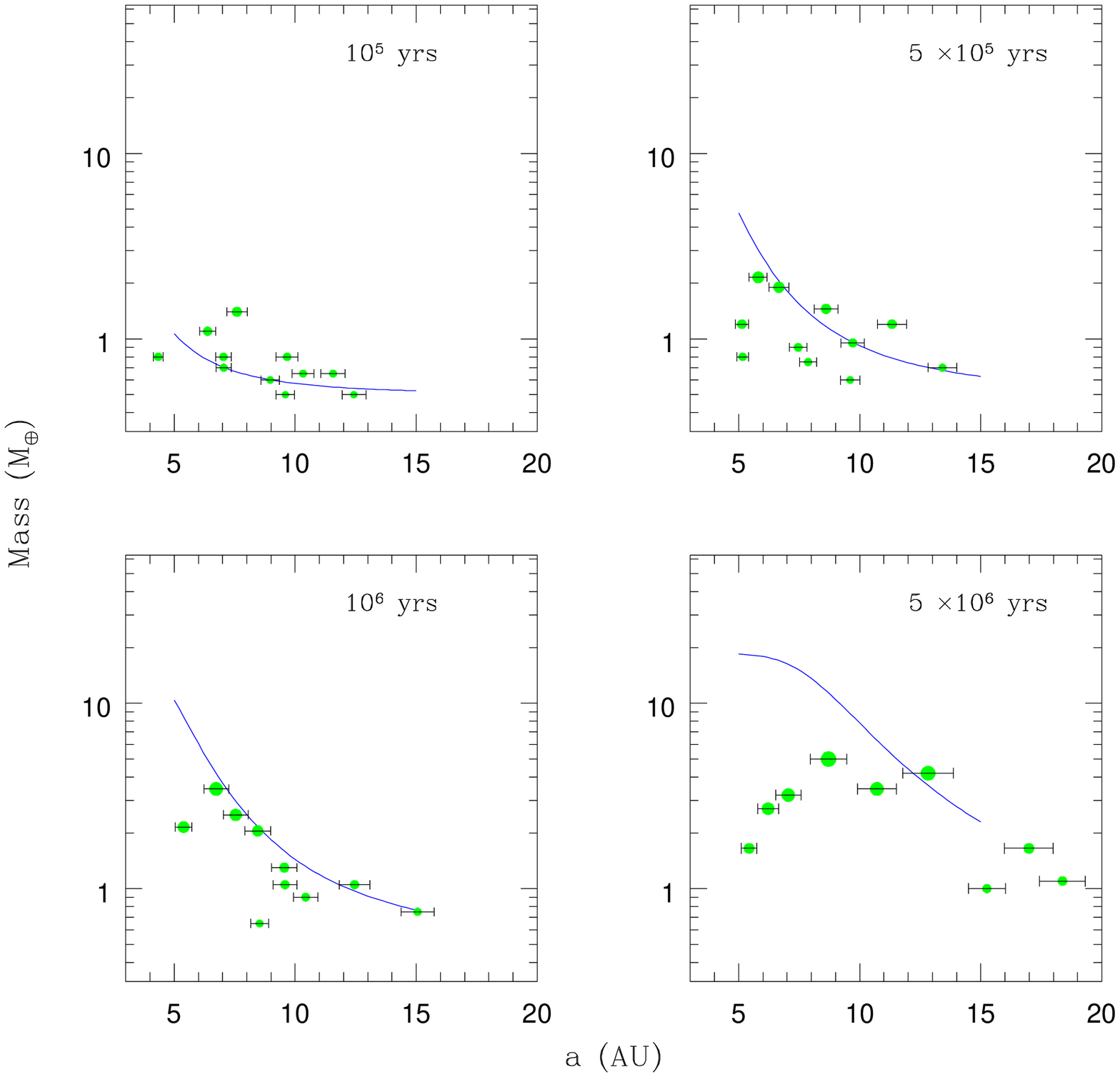}
\caption{Snapshots of the protoplanet masses and semimajor axes in Run
D at different times, together with the model-predicted mass.  The
horizontal bars on the protoplanets are of length 10 $r_{\rm H}$. }
\label{d_ti_10km_massonly}
\end{center}
\end{figure}

\begin{figure}[p]
\begin{center}
\includegraphics[width=5.0in]{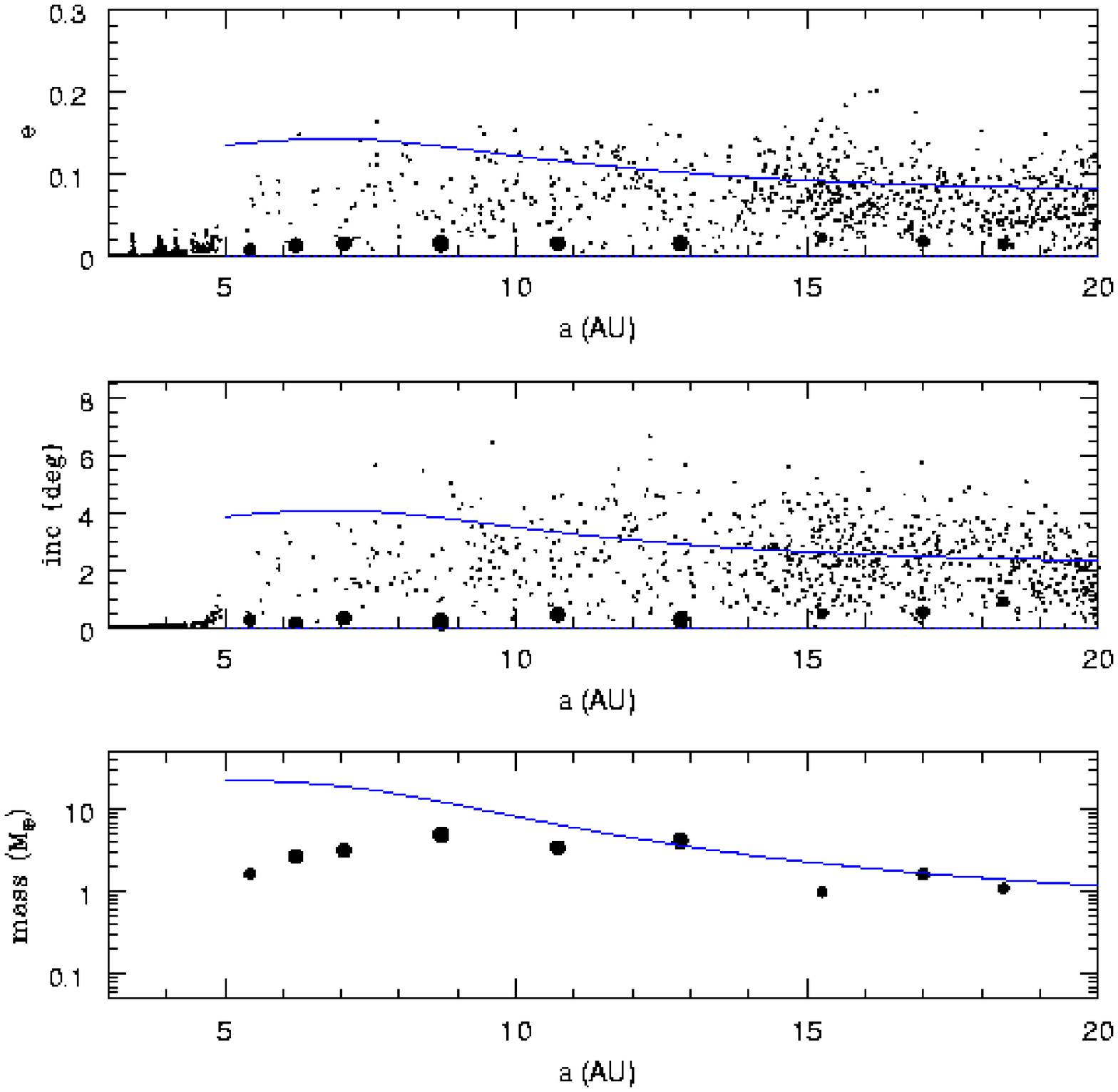}
\caption{The state of Run D at its endpoint of 5 Myrs, showing
eccentricities (top), inclinations (middle) and protoplanet masses
(bottom) together with their predicted values.}
\label{d_ti_10km_endstate}
\end{center}
\end{figure}

The length of this run is 5 Myrs; snapshots at 0.1, 0.5, 1 and 5 Myrs
are shown in Fig. \ref{d_ti_10km_massonly}, and the endstate is shown
in more detail in Fig. \ref{d_ti_10km_endstate}.  The run produces a
largest body of 5 M$_{\oplus}$ just inside 9 AU by 5 Myrs, and a total
of four bodies that are more massive than 3 M$_{\oplus}$.  The initial
growth rate is again faster than predicted, but the masses produced
once the wave of growth has swept past, inside 9 AU at 5 Myrs, are
lower than the theoretical value by a factor of several.
Fig. \ref{d_ti_10km_endstate} shows only a small number of
planetesimals remaining among the protoplanets interior to the growth
front, confirming that accretion there has essentially concluded.

\section{Conclusions}
\label{discussion}

Runaway growth allows very short formation times, but only in the
early stages of planetesimal accretion; there is a transition to the
self-limiting oligarchic growth mode when the largest bodies are still
orders of magnitude below an Earth mass.  The timescale of oligarchic
growth thus dominates over that of runaway growth, and we use the
former alone to obtain a global picture of planet formation throughout
a protoplanetary disk.  In the terrestrial region, accretion
efficiency is high, and oligarchic growth alone is not required to
form final bodies of terrestrial-planet masses.  Simulations show that
a late stage of impacts among protoplanets of up to perhaps Mars mass
is readily able to produce a final system with planet masses and
spacings similar to the present-day inner Solar System (eg. Chambers
and Wetherill 1998).  A significant planetesimal population is not
necessary to facilitate late-stage accretion, though it may be needed
to reproduce (via dynamical friction) the low eccentricities of the
terrestrial planets.

In the giant planet region, however, lower accretion efficiency and
the necessity of forming large bodies before the removal of the
nebular gas (after $\la 10$ Myrs) mean that direct oligarchic growth
likely has to be relied on to account for most of the accretion.  In
our semi-analytic model, the outward-sweeping front of oligarchic
growth produces 10 M$_{\oplus}$ bodies in a protoplanetary disk with
ten times the mass of the Hayashi (1981) minimum-mass model in less
than a million years.  These first appear near the snow line, where
the surface density enhancement gives accretion a head start and
increases the maximum attainable masses.  Thus, the snow line may
indeed play a role in triggering the formation of giant planets, as
has been previously suggested (Morfill 1985).

\begin{figure}[p]
\begin{center}
\includegraphics[width=5.0in]{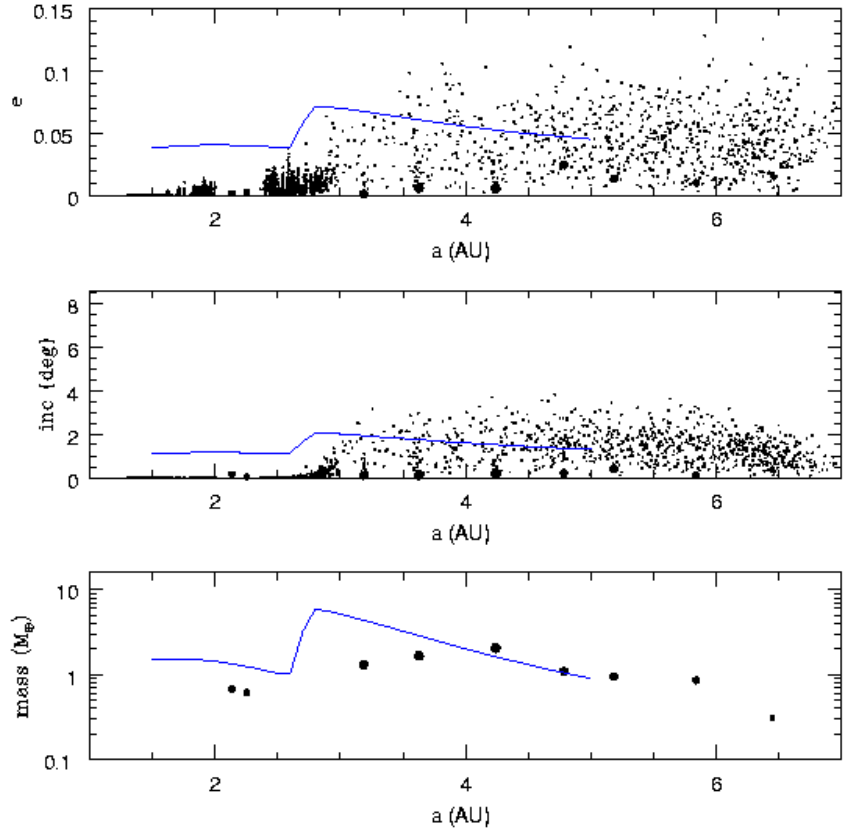}
\caption{A more detailed snapshot of the state of Run A at its
endpoint of 1 Myr, showing also the planetesimals.  Eccentricities
(top), inclinations (middle), and protoplanet masses are plotted
versus semimajor axis, with superimposed curves showing their
model-predicted values.}
\label{1e6_paper}
\end{center}
\end{figure}

\begin{figure}[p]
\begin{center}
\includegraphics[width=5.0in]{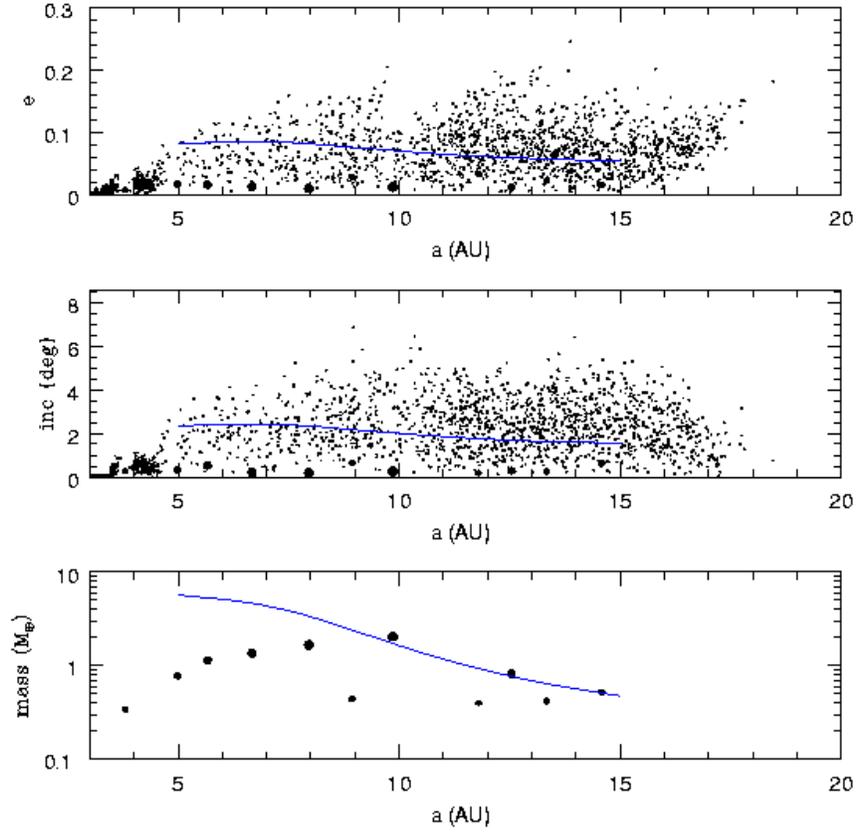}
\caption{A more detailed snapshot of the state of Run B at its
endpoint of 10 Myrs, showing eccentricities (top), inclinations
(middle) and protoplanet masses (bottom) together with their predicted values.}
\label{1e7_log}
\end{center}
\end{figure}

\begin{figure}[p]
\begin{center}
\includegraphics[width=5.0in]{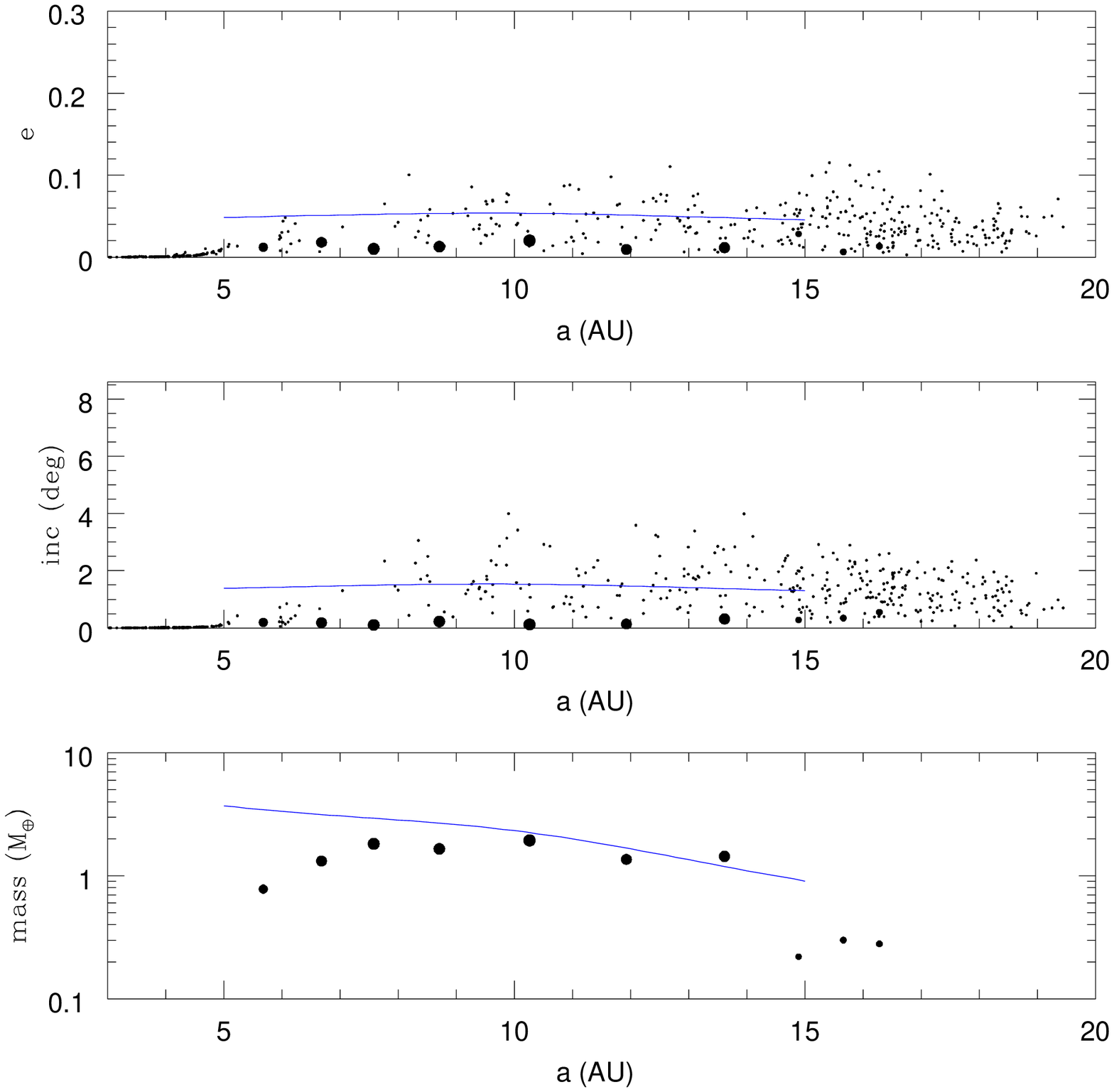}
\caption{The state of Run C at its endpoint of 10 Myrs, showing
eccentricities (top), inclinations (middle) and protoplanet masses
(bottom) together with their predicted values.}
\label{1km_10myrs}
\end{center}
\end{figure}

The numerical simulations performed here agree reasonably well with
the model insofar as the timing of the initial oligarchic growth front
is concerned.  However, accretion stalls earlier than predicted, and
the largest final masses produced typically fall short of the
theoretical values by factors of several.  This can be attributed in
part to a simulation edge effect, namely, the spreading of the
simulated planetesimal disk beyond its initial radial extent, which
causes the disk surface density to decrease faster than the model
(which does not incorporate this effect) predicts; the effect is
strongest at the inner edge of the disk.  Another edge effect is that
the innermost few protoplanets act more like isolated bodies in a
planetesimal disk, because there are fewer neighbouring stirred
regions which overlap theirs.  Consequently, they tend to open a gap
in the planetesimals, thus putting a premature end to their growth
(Rafikov 2001).  Gap formation about the innermost protoplanets is
evident in Figs. \ref{1e6_paper}, \ref{1e7_log} and \ref{1km_10myrs}.
In order to make the inner edge less artificial, future simulations
ought to include additional larger protoplanets interior to the
planetesimal disk, to play the role of the ``endproducts'' in the
region where oligarchic growth has already reached completion.
However, given that protoplanets interior to the snow line are
expected to reach significantly smaller final masses, edge effects
like those in our simulations may not be completely unrealistic.  In
any case, it appears that the semi-analytic estimate we have developed
here for the oligarchic growth endproduct masses is an upper limit.
 
Overall, then, it is still somewhat of a challenge to understand how
(potential) gas giant cores can form in a protoplanetary disk.  The formation of an extended gas
atmosphere on larger protoplanets, not modeled here, will certainly
help in sweeping up planetesimals.  It may simply be that the
protoplanetary disks which produce giant planet-bearing systems really
are quite massive, perhaps in excess of ten times the minimum mass.
 However, one can only push the disk mass so far before one comes up
against observational limits.  Stevenson and Lunine (1988) develop a
model in which diffusive redistribution of water vapor from the inner
part of the nebula leads to a large local density enhancement near the
snow line, so that the jump in solids density there has an additional
spike superimposed on it.  
Another possible scenario is that runaway gas accretion already takes place
at smaller core masses, perhaps as little as 1 M$_{\oplus}$.
This would be facilitated by low grain opacites in the accreting
atmosphere (eg. Inaba and Wetherill 2001).  Yet another possibility is
that the window of opportunity for gas giant formation is larger than
normally assumed.  Kokubo and Ida (2002) adopt $10^8$ rather than
$10^7$ years as the gas lifetime, pointing out that the timescale for
the {\it complete} dispersal of gas may approach this larger value
(Thi et al 2001).  However, as the gas density drops, so will the
planetesimal accretion rate.  Also, it is unclear how much the gas
nebula can be depleted before core accretion stops being viable.

In 10 Myrs, the (optimistic) predicted growth front only produces 10
M$_{\oplus}$ bodies interior to about 10 AU, even in very massive
protoplanetary disks.  Thus in the case of our Solar System, although
the solid cores of Jupiter and Saturn can be more or less accounted
for, it seems that Uranus and Neptune are out of luck as far as in
situ formation during the lifetime of (most of) the gas is concerned.
Accounting for the ice giants' gas content is a problem, since the
protoplanets which are predicted to have formed in the trans-Saturnian
region by the time the gas disperses are too small to have acquired
appreciable atmospheres.  The most fundamental difficulty, however, is
that of the post-gas growth timescales.  If random velocities in the
planetesimal disk are not damped at all, then the formation of Uranus-
and Neptune-mass planets is not possible on {\it any} timescale.  If
collisions among planetesimals are an effective dissipational
mechanism, then such planets could conceivably form, though it could
well take of order a billion years, during which time the planetesimal
disk has to maintain a sufficiently high optical depth to keep the
collisions going.

A model in which the ice giants shared the same birthplace as Jupiter
and Saturn, only to be scattered outward when Jupiter accreted its
massive gas envelope, provides an alternative (Thommes, Duncan and
Levison 1999, 2002).  Of course, this scenario is predicated on the
ability to form multiple $\ga$ 10 M$_{\oplus}$ bodies in the
Jupiter-Saturn zone, notwithstanding the above difficulties, while at
the same time leaving some of them without massive gas envelopes at
the time when the nebular gas disappears.  The simulations performed
here provide some support for this scenario, in the sense that they
show several comparable-mass largest protoplanets occupying the gas
giant region after several million years of evolution.  Run D, with
the largest initial disk mass, looks the most promising.  However, the
protoplanets produced still fall significantly short of Uranus and
Neptune's core masses.  Run D's protoplanet masses at 10 Myrs would
need to be scaled up by a factor of about three; then the largest
protoplanets would comprise a system not dissimilar to the initial
conditions used in Thommes, Duncan and Levison (1999, 2002)

Orbital evolution through disk tides (eg. Ward 1997) is not considered
in this analysis.  Fast Type I inward migration, predicted to be
strongest for bodies with masses of order 10 - 100 M$_{\oplus}$,
constitutes a major potential problem for any model of giant planet
formation, except ones which invoke direct, unnucleated collapse from
a gas disk instability (eg. Boss 1998).  However, at present it is
still quite uncertain just how serious this problem is.  For instance,
it has been shown that the nature of the migration depends sensitively
on the assumed disk boundary conditions (Tanaka, Takeuchi and Ward
2002).  Also, Type I migration may be halted at a smaller mass than is
commonly assumed (eg. Rafikov 2002).

In the limit of no Type I migration, the investigation presented here
suggests the following overall picture of the post-runaway planet
formation process: During the first ten million years or less, while
the nebular gas is still present, a relatively rapid front of
oligarchic growth sweeps outward through the disk.  By the time the
gas disperses, however, this wave has only reached what corresponds to
the Jupiter-Saturn region; very little accretion takes place during
this time in the trans-Saturnian region, thus the planetary system is
initially quite compact, with a Kuiper belt starting at $\sim$ 10 AU.
Bodies large enough to be potential gas giant cores form only in an
annulus encompassing roughly the outer half of the (proto-) planetary
system's radial extent.  Once the gas disappears, the
outward-expanding wave of growth becomes far slower and the subsequent
formation of giant planets (of necessity gas-poor ice giants like
Uranus and Neptune) becomes difficult, perhaps impossible.  However,
if gas giants are able to form before the removal of the gas, a likely
by-product will be the scattering of any remaining giant protoplanets
that missed out on acquiring a massive gas envelope.  This could
ultimately produce an outer planetary system with widely-spaced,
circular giant planet orbits like those of the present-day Solar
System, while requiring little or no accretion to take place after the
gas is gone.

\acknowledgements 

\noindent {\it Acknowledgements:} This work is supported by the Center
for Integrative Planetary Science (EWT), NASA's {\it Origins of Solar
Systems} (EWT, HFL), {\it Planetary Geology \& Geophysics} and {\it
Exobiology} (HFL) programs, and by Canada's National Science and
Engineering Research Council (MJD).  We would like to thank the
referees, Satoshi Inaba and Eiichiro Kokubo, for valuable suggestions
which helped us to improve the manuscript.  Also, we would like to
thank Glen Stewart and Andrew Youdin for helpful discussions.


\clearpage

\begin{figure}[p]
\begin{center}
\includegraphics[width=5.0in]{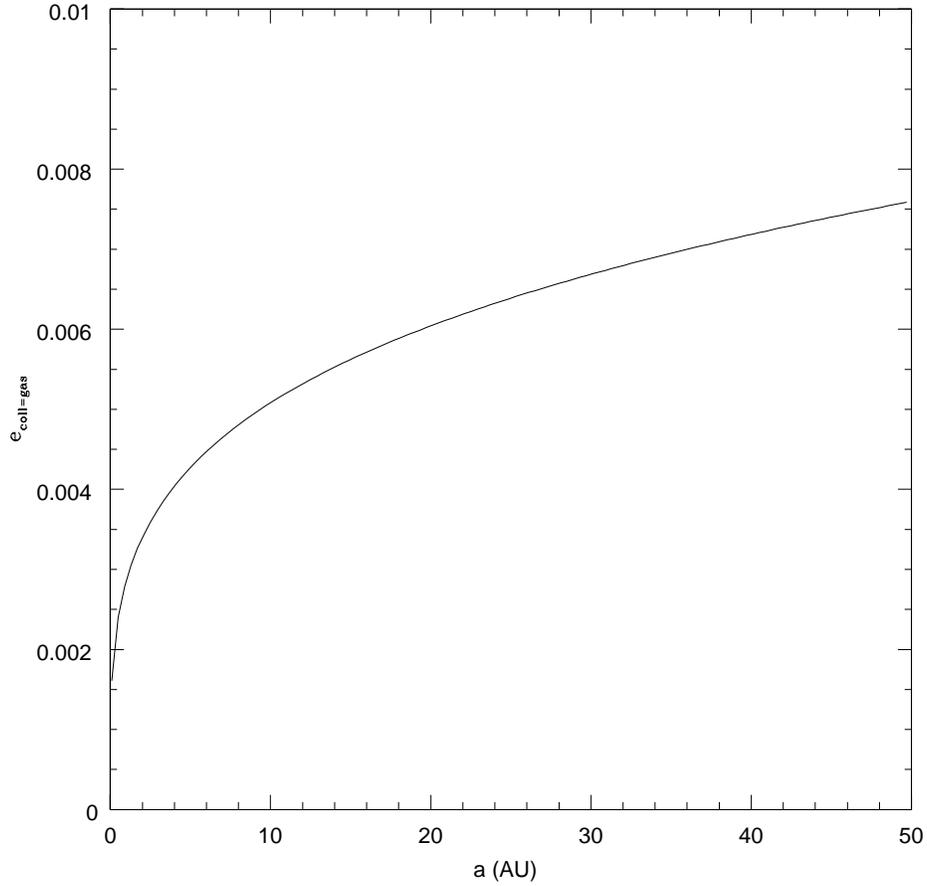}
\caption{For a Hayashi-profile disk, the eccentricity $e_{\rm coll=gas}$
at which the timescale for random velocity damping by gas drag comes
to be equal to the planetesimal-planetesimal collision time.  At
higher eccentricities, the gas timescale is shorter.}
\label{collgas} 
\end{center}
\end{figure}

\clearpage

\begin{table}[p]
\caption{Simulation parameters}
\label{simtable}
\begin{center}
\footnotesize
\begin{tabular}{l l l l l}
\hline 
Quantity & Run A & Run B & Run C & Run D\\ 
\hline
initial solids
surface density (g/cm$^2$) & $150 \left ( \frac{a}{\mbox{1 AU}} \right
)^{-3/2}$ & $250 \left ( \frac{a}{\mbox{1 AU}} \right )^{-2}$ & $250
\left ( \frac{a}{\mbox{1 AU}} \right )^{-2}$ & $225 \left (
\frac{a}{\mbox{1 AU}} \right )^{-3/2} $\\ 
gas volume density
($10^{-9}$g/cm$^3)$ & $7\left (
\frac{a}{\mbox{1 AU}} \right )^{-11/4}$  & $11.5 \left (
\frac{a}{\mbox{1 AU}} \right )^{-13/4}$ & $11.5 \left (
\frac{a}{\mbox{1 AU}} \right )^{-13/4}$ &$10.5\left (
\frac{a}{\mbox{1 AU}} \right )^{-11/4}$ \\
initial radial extent & 1.5 - 5 & 5 - 15 & 5 - 15 & 5 - 15\\
initial protoplanet mass (M$_{\oplus}$) & 0.1 & 0.2 & 0.2 & 0.5\\ 
initial protoplanet number & 22 & 16 & 16 & 12\\
planetesimal mass & 0.01 & 0.02 & 0.02 & 0.05\\
initial planetesimal number & 4886 & 3233 & 3233 & 3446\\
planetesimal gas drag radius & 10 km & 10 km & 1 km & 10 km\\
\hline
\end{tabular}
\normalsize
\end{center}
\end{table}

\end{document}